\documentclass[12pt,a4paper]{article}
\usepackage[utf8]{inputenc}
\usepackage[english]{babel}
\usepackage{amsmath}
\usepackage{amsfonts}
\usepackage{amssymb}
\usepackage{graphicx}
\usepackage[section]{placeins}
\FloatBarrier
\usepackage{flafter}
\usepackage{cancel}
\usepackage{color}
\usepackage{slashed, epsf, latexsym}
\usepackage{epsfig}
\usepackage{graphics}
\usepackage{jheppub}

\begin{document}
\preprint{}
	\title{Comparison between fluid-gravity and membrane-gravity dualities for Einstein-Maxwell system}
	\author[]{Milan Patra}
	\affiliation[]{National Institute of Science Education and Research, HBNI, Bhubaneshwar 752050, Odisha, India}
	\emailAdd{milan.patra@niser.ac.in}

\abstract{Derivative expansion and large-$D$ expansion are two perturbation techniques, which are used to generate dynamical black-brane solutions to Einstein's equations in presence of negative cosmological constant. In this note we have compared these two techniques and established the equivalence of the gravity solutions generated by these two different techniques in appropriate regime of parameter space up to first non-trivial order in both the perturbation parameters for Einstein-Maxwell systems, generalizing the earlier works of \cite{Bhattacharyya2019leading,Bhattacharyya2019sunleading} for non-charged systems. An one-to-one map between dynamical black-brane geometry and AdS space, which also exists at finite number of dimensions, has also been established.}
\maketitle
\section{Introduction :}
Over the last few years, it has been demonstrated that Black hole dynamics at large number of space-time dimensions (denoted as $D$ here) is dual to the dynamics of a co-dimension one non-gravitational membrane propagating in a background that solves Einstein's equations. Emparan, Suzuki and Tanabe \cite{Emparan:2013moa,QNM:Emparan,Emparan:2013xia,Emparan:2013oza,Emparan:2014cia,Emparan:2014jca} have observed that at large number of space-time dimensions the black holes have two effective length scales; one is the radius of the horizon $r_H$ and  the other is the thickness of the region over which the gravitational force is non trivial. Beyond this region the space-time, to a good approximation, could be described the asymptotic geometry itself (for our purpose this is going to be the AdS space). It turns out that in the limit of large $D$, this thickness is of the order of $\mathcal{O}\left(\frac{r_H}{D}\right)$ around the horizon and within this region Einstein equations could simply be solved in a perturbation in inverse powers of dimension.\\
 Within this perturbation scheme, the effective dynamics of these black hole horizons can be described by a co-dimension one massive membrane with dynamical degrees of freedom as charge density, shape and a divergence-free velocity field moving in the background space. This duality has been studied for both asymptotically flat and AdS/dS background and also for Einstein-Maxwell systems in \cite{membrane,Chmembrane,yogesh1,arbBack,Kundu2018chargeAdsds,Bhattacharyya20182nd}.\\
\par 
On the other hand `derivative expansion', which is a perturbation technique in boundary derivative expansion \cite{nonlinfluid,arbitrarydim,Rangamani:2009xk,Banerjee:2008th,Erdmenger:2008rm,Haack:2008cp,Bhattacharyya:2007vs,Banerjee:2012iz,Bhattacharyya:2012xi,Hubeny:2011hd,Bredberg:2011xw}, generates dynamical black-brane solutions of Einstein's equations in the bulk with negative cosmological constant. These dynamical black brane solutions are dual to an arbitrary fluid dynamical solutions in the boundary. In other words, for every solution to the relativistic generalization of Navier-Stokes' equation in the boundary, one can construct an asymptotically AdS black hole type solution equivalent to a boundary fluid flow. These solutions thus generated are characterized by a local temperature field, a unit normalized velocity field and charge density living on the boundary.\\
\par
The questions that we would like to address in this note are the following.
\begin{itemize}
\item  Is there any interconnection between these two perturbation techniques
\item  Can we apply these two techniques simultaneously in any regime of parameter space
\item Are the solutions generated by these two techniques equivalent in any regime of parameter space?
\end{itemize}
 These questions have been answered in \cite{Bhattacharyya2019leading,Bhattacharyya2019sunleading} for pure gravity systems and here we shall extend their work to  Einstein-Maxwell systems. We shall show that the gravity solutions generated by these two perturbation techniques are equivalent also for Einstein-Maxwell systems in an appropriate regime of parameter space.\par
On physical grounds this equivalence is not surprising. Since we can use the same space-time geometry as the starting point for both the techniques and since given  a starting point both the techniques generate the solutions uniquely, it follows that  in the regime where both the techniques are applicable, the solutions should be same. But this is not at all manifest and it involves a series of intricate gauge and variable transformations. Just in \cite{Bhattacharyya2019leading,Bhattacharyya2019sunleading} here also the heart of the paper lies in these set of subtle calculations.\par

\subsection{Strategy}
In this subsection we will discuss briefly the procedure we have used to show the equivalence of the two gravity solutions and refer to \citep{Bhattacharyya2019leading,Bhattacharyya2019sunleading} for any logical discussion and proof. As we know the metric $W_{AB}$ generated in large-$D$ technique are written in a split form, a background metric $\bar{W}_{AB}$ and $W^{(rest)}_{AB}$. The metric $\bar{W}_{AB}$ is the metric of the asymptotic geometry (in our case it is pure AdS metric) and $W^{(rest)}_{AB}$ is written in a way such that contraction of a certain null geodesic $O^A\partial_A$ (not affinely parametrized) with it is zero to all order. But the hydrodynamic metric is not written in such split form. So it is obvious that to compare the solutions generated by these two techniques the first step should be to split the hydrodynamic metric into background and `rest'. We will do it by the following steps\\

At first we determine an affinely parametrized null geodesic field $\bar{O}^{A}\partial_A$, which passes through the event horizon of the full space-time. Then we pick up a coordinate system $Y^{A}\equiv \{\rho,y^{\mu}\}$, where the background of the hydrodynamic metric can be written in the following form
\begin{equation}
ds^2_{background}=\bar{G}_{AB}dY^AdY^B=\frac{d\rho^2}{\rho^2}+\rho^2\eta_{\mu\nu}dy^{\mu}dy^{\nu}
\end{equation}
And determine the mapping function $f^A(X)$ that relates $Y^A$ to the $X^A\equiv \{r,x^{\mu}\}$ (in which the hydrodynamic metric is written) coordinates by the following equation
\begin{equation}
\bar{O}^A{\cal G}_{AB}|_{\{X\}}=\bar{O}^A\frac{\partial f^C}{\partial X^A}\frac{\partial f^{C'}}{\partial X^B}\bar{G}_{CC^{\prime}}|_{\{X\}}
\end{equation}
where ${\cal G}_{AB}$ is the full metric written in $X^A$ coordinates and the subscripts $\{X\}$ refer to the fact that all the terms in the left and right are calculated in $X^A$ coordinates.\par
However it cannot fix $f^A$ completely and we require some conformal type symmetry on the background metric to fix it.\par 
After determining the mapping function we can split the hydrodynamic metric into background plus `rest'. Then we take the large-$D$ limit of the hydrodynamic metric.\\

The next step is to write the large-$D$ data in terms of fluid data. The metric and gauge field generated in large-$D$ technique are written in terms of a smooth function $\psi$ (such that $\psi^{-D}$ is a harmonic function w..r.t the background), a charge field $\widetilde{Q}$ and a non-affinely parametrized null geodesic $O_A$. It will turn out that $O_A$ is related to $\bar{O}_A$ (determined from the hydrodynamic metric) by an overall normalization constant.\par 
At first we determine $\psi$ and then $O_A$ and $\widetilde{Q}$ in terms of fluid data. After that we substitute these expressions in the large-$D$ metric and gauge field and write those in terms of fluid data. Then it is easy to check that the metric and gauge field in large-$D$ side matches with those in fluid side up to appropriate orders in both the perturbation parameters.\\

But the hydrodynamic data and large-$D$ data cannot be chosen arbitrarily, they have to satisfy some constraint equations, named as fluid equations and membrane equations respectively for the metric and gauge field to be a solution of Einstein's equations. So to show the equivalence of the gravity solutions generated by these two techniques the final step would be to show the equivalence of these constraint equations.\\

 The organisation of this note is as follows.\\
In section \S \ref{sec:Intro} we review the hydrodynamic metric and gauge field up to first order in derivative in arbitrary dimensions. In section \S \ref{largeDmetricgauge} we have noted the gravity solutions for Einstein-Maxwell systems in large-$D$ technique. In section \S \ref{sec:gr_solution_comp} we have rewritten the large-$D$ data in terms of fluid data and compared the two gravity solutions. And finally in section \S \ref{conclusion} we have concluded.

\section{Review of Hydrodynamics from charged black-branes in arbitrary dimensions :}\label{sec:Intro}
In this section we will review the work on fluid-gravity correspondence for charged black-branes \cite{Banerjee:2008th,Erdmenger:2008rm,articleyee} by determining the metric and gauge field dual to charged fluid configuration up to first order in boundary derivative expansion for all $D\geq 3$. The results of this section were previously recorded in \cite{articleyee} in a bit different language. As our lagrangian and notations are slightly different from the authors of \cite{articleyee}, we will redo everything with our lagrangian and notations. \par
We start with the $D$ dimensional action\footnote{We will use the Latin indices $\{M, N, \cdots\}$ to denote the bulk indices while the Greek indices $\{\mu, \nu, \cdots\}$ refer to the boundary indices. And the $\{\mu,\nu,\cdots\}$ indices are raised and lowered by the Minkowaski metric $\eta^{\mu\nu}$. And for our case $\lambda=-1$.}
\begin{equation}
 \begin{split}
&\mathcal{\mathbf{S}}=\frac{1}{16\pi G_D}\int~d^Dx~\sqrt{-\text{g}}\left[R-2\Lambda-\frac{F_{MN}F^{MN}}{4}\right]\\\
&\text{with negative cosmological constant}~\Lambda=\frac{\left(D-1\right)\left(D-2\right)}{2}\lambda
\end{split}
\end{equation}
By varying the above action we will get the $D$ dimensional Einstein-Maxwell equations with negative cosmological constant 
\begin{equation}\label{EM_equation}
 \begin{split}
 & R_{AB}-\frac{1}{2}R~g_{AB}-\frac{(D-1)(D-2)}{2}g_{AB}+\frac{1}{2}~F_{AC}F^{C}_{~B}+\frac{1}{8}~g_{AB}F_{CD}F^{CD}=0\\
 &\nabla_{B}F^{AB}=0\Rightarrow \frac{1}{\sqrt{-g}}\partial_{B}\left(\sqrt{-g}~F^{AB}\right)=0 
 \end{split}
\end{equation}
where $g_{AB}$ is the $D$ dimensional metric tensor and $F_{AB}=\partial_{A}A_{B}-\partial_{B}A_{A}$.\par
We know that these equations \eqref{EM_equation} admit an AdS-Reisner-Nordstrom `boosted black-brane solutions', which we write in ingoing Eddington-Finkelstein coordinates as
\begin{equation}\label{defblack_brane}
  \begin{split}
   d&s^2 =-2u_{\mu}dx^{\mu}dr-r^2V\left(r, m, Q\right)u_{\mu}u_{\nu}dx^{\mu}dx^{\nu}+r^2P_{\mu\nu}dx^{\mu}dx^{\nu}\\
 &A=\frac{\sqrt{3}~Q}{2r^{D-3}}u_{\mu}dx^{\mu}\\
  \end{split}
  \end{equation}
  with
 \begin{equation}
  \begin{split}
   &V(r,m,Q)=1-\frac{m}{r^{D-1}}+\frac{1}{4}~\frac{3(D-3)}{2(D-2)}\frac{Q^2}{r^{2(D-2)}}~,~~~P^{\mu\nu}=\eta^{\mu\nu}+u^{\mu}u^{\nu}\\
   &u^v=\frac{1}{\sqrt{1-\beta^2}}~,~u^i=\frac{\beta^i}{\sqrt{1-\beta^2}}~,~\beta^2=\beta_i\beta^i\\
  \end{split}
  \end{equation}
 Like in the metric described above we will use coordinates $X^{A}\equiv \underbrace{\{r,v,x^{i}\}}_{D}$ for our bulk spaces, on the other hand coordinates $x^{\mu}\equiv \underbrace{\{v,x^i\}}_{D-1}$ parametrize our boundary and $r$ is the radial coordinate.\\
 \par
Now we allow the temperature, velocity and charge field in the black-brane metric \eqref{defblack_brane} to vary slowly in the boundary coordinates and determine the metric and gauge field in boundary derivative expansion. To start we will take the ansatz as
\begin{equation}
\begin{split}
&g_{MN}=g^{(0)}_{MN}+g^{(1)}_{MN}+g^{(2)}_{MN}+\cdots\\
&A_{N}=A^{(0)}_{N}+A^{(1)}_{N}+A^{(2)}_{N}+\cdots\\
\end{split}
\end{equation}
 where the leading order ansatz $g^{(0)}_{MN}$ and $A^{(0)}_{N}$ are given by 
 \begin{equation}
 \begin{split}
 &g^{(0)}=-2u_{\mu}dx^{\mu}dr-r^2V(r, m, Q)u_{\mu}u_{\nu}dx^{\mu}dx^{\nu}+r^2P_{\mu\nu}dx^{\mu}dx^{\nu}\\
 &A^{(0)}=\frac{\sqrt{3}~Q}{2r^{D-3}}u_{\mu}dx^{\mu}\\
 \end{split}
 \end{equation}
and $g^{(k)}_{MN}$, $A^{(k)}_N$, which are corrections to the bulk metric and gauge field, are determined by solving Einstein-Maxwell equations order by order in derivative expansion. To solve these equations by our perturbation technique it is useful to work in a particular gauge. Following\cite{Banerjee:2008th} we work in the following gauge
\begin{equation}
 \begin{split}
 g_{rr}=0~,~g_{r\mu}=-u_{\mu}~,~\text{Tr}\left[\left(g^{(0)}\right)^{-1}g^{(k)}\right]=0~,~A_r=0
 \end{split}
 \end{equation}
Here one should note that in the relativistic case the energy flow between different fluid elements also leads to transport of mass and momentum. Hence we cannot define the velocity field uniquely unless we work in a particular frame. In this context it is useful to work in Landau frame defined by $u_{\mu}T^{\mu\nu}_{(k)}=0$, where $T^{\mu\nu}_{(k)}$ is the $\text{k}^{th}$ order stress tensor with $k\ge 1$ and in the proper frame of a fluid element the longitudinal component of the stress tensor to the fluid velocity give the local energy density in the fluid.\par
 In this section our goal is to find out the metric and gauge field up to first order in derivative expansion. To implement our perturbation technique we set our velocity field $u^{\mu}$ to be $\{1,0,0,\cdots\}$ by a boundary Lorentz transformation at a boundary point $x^{\mu}$ and solve these equations about this special point. Since our perturbation procedure is ultra-local we can easily write the result thus obtained in covariant form with respect to the boundary metric. Our velocity, temperature and charge field expanded in taylor series about this special point $x^{\mu}$ in terms of boundary derivatives as
 \begin{equation}
 \begin{split}
 &\beta_i=x^{\mu}\partial_{\mu}\beta^{(0)}_i+\cdots\\
 &m=m^{(0)}+x^{\mu}\partial_{\mu}m^{(0)}+\cdots\\
 &Q=Q^{(0)}+x^{\mu}\partial_{\mu}Q^{(0)}+\cdots\\
 \end{split}
 \end{equation}
Using the expressions written above if we expand the $0^{th}$ order ansatz up to first order in derivative we have 
 \begin{equation}\label{zeroth_expansion}
 \begin{split}
 ds^2_{(0)}=&2~dv~dr-r^2V^{(0)}(r)~dv^2+r^2dx^i~dx_i\\
 &-2r^2\left(1-V^{(0)}(r)\right)x^{\mu}\partial_{\mu}\beta^{(0)}_i~dx^i~dv-2x^{\mu}\partial_{\mu}\beta^{(0)}_i~dx^idr\\
 &-\left(-\frac{x^{\mu}\partial_{\mu}m^{(0)}}{r^{D-3}}+\frac{1}{4}\frac{3(D-3)}{2(D-2)}\frac{2Q^{(0)}x^{\mu}\partial_{\mu}Q^{(0)}}{r^{2(D-3)}}\right)dv^2\\
 A=&-\frac{\sqrt{3}}{2r^{D-3}}\left[\left(Q^{(0)}+x^{\mu}\partial_{\mu}Q^{(0)}\right)dv-Q^{(0)}x^{\mu}\partial_{\mu}\beta^{(0)}_idx^{i}\right]\\
 \text{with}~~~&V^{(0)}(r)=1-\frac{m^{(0)}}{r^{D-1}}+\frac{1}{4}~\frac{3(D-3)}{2(D-2)}\frac{{Q^{(0)}}^2}{r^{2(D-2)}}
\end{split}
\end{equation}
 Obviously this metric and gauge field of equation \eqref{zeroth_expansion} will not solve the Einstein-Maxwell's equations up to the order we are interested. We need to add corrections containing first order derivatives to solve these equations. While solving these equations we find that the bulk Einstein-Maxwell equations decompose into constraint equations and dynamical equations. The constraint equations are equivalent to the conservation of boundary stress energy-momentum tensor and conservation of boundary current density. On the other hand the dynamical equations are inhomogeneous differential equations on the unknown parameters added to the metric and gauge field. By solving these differential equations and imposing regularity at the future event horizon and appropriate fall off at infinity, we can determine the unknown parameters. Thus we get an unique solution to the Einstein-Maxwell equations parametrized by the charge, temperature and $(D-1)$ dimensional velocity fields.
 \par Since the background metric have the $SO(D-2)$ symmetry, the Einstein-Maxwell equations will split into scalar, vector, and traceless symmetric two tensor sector and we can solve each sector separately. 
\subsection{Scalars at first order}
The scalar components of the added correction of the first order metric and gauge field are parametrized by $h_1(r), k_1(r)$ and $w_1(r)$ and we can write these as
\begin{equation}
 \begin{split}
&g^{(1)}_{vv}(r)=\frac{k_1(r)}{r^2}\\
&g^{(1)}_{vr}(r)=-\frac{\left(D-2\right)}{2}h_1(r)\\
&\sum_i~g^{(1)}_{ii}(r)=(D-2)~r^2~h_1(r)\\
&A^{(1)}_v(r)=-\frac{\sqrt{3}~w_1(r)}{2~r^{D-3}}\\
\end{split}
\end{equation}
Here one should note that the metric corrections $g^{(1)}_{ii}$ and $g^{(1)}_{vr}$ are related to each other by the gauge choice $\text{Tr}\left[\left(g^{(0)}\right)^{-1}g^{(1)}\right]=0$. At first we determine the constraint equations. These equations are determined by taking dot product of Einstein-Maxwell equations with the vector dual to the one form $dr$. We have these equations as follows
\begin{equation}
\begin{split}
 \frac{\partial_vm^{(0)}}{m^{(0)}}+\frac{(D-1)}{(D-2)}\partial_i\beta^{(0)}_i=0\\
\end{split}
\end{equation}
 which is identical to the conservation of the stress tensor in the boundary
 \begin{equation}
 \begin{split}
 \partial_{\mu}T^{\mu\nu}_{(0)}=0
  \end{split}
\end{equation} 
The other constraint equation is given by
\begin{equation}
\begin{split}
 \frac{\partial_vQ^{(0)}}{Q^{(0)}}+\partial_i\beta^{(0)}_i=0\\
\end{split}
\end{equation} 
which is equivalent to the conservation of boundary current density.
\begin{equation}
 \begin{split}
 \partial_{\mu}J^{\mu}_{(0)}=0\\
 \end{split}
\end{equation}
In the scalar sector we have 6 differential equations, the $rr, vv, rv$ component of the Einstein tensor along with the trace over the boundary spatial part and the $r$ and $v$ components of the Maxwell equations. Among these 6 equations we have to use only 3 equations to determine the three unknown parameters $h_1, k_1~\text{and}~w_1$. The solutions thus obtained should satisfy the rest equations. Solving these equations and demanding the appropriate normalizability at infinity we have the following solutions  
\begin{equation}
 \begin{split}
  &h_1(r)=0\\
  &w_1(r)=0\\
  &k_1(r)=2~r^3\frac{\partial_i\beta^{(0)}_i}{D-2}
 \end{split}
\end{equation}
 Finally we have the first order metric and gauge field in the following form 
 \begin{equation}
 \begin{split}
&g^{(1)}_{vv}(r)=2r\left(\frac{\partial_i\beta^{(0)}_i}{D-2}\right)\\
&g^{(1)}_{vr}(r)=0\\
&\sum_i~g^{(1)}_{ii}(r)=0\\
&A^{(1)}_v(r)=0\\
\end{split}
\end{equation}
\subsection{Vectors at first order}
 The vector components of first order metric and gauge field are parametrized by $g^{(1)}_i(r)$ and $j^{(1)}_i(r)$ as
\begin{equation}\label{correct_vector}
\begin{split}
& g^{(1)}_{vi}(r)=r^2\left(1-V(r)\right)j^{(1)}_i(r)\\
 &A^{(1)}_i(r)=-\frac{\sqrt{3}Q^{(0)}}{2r^{D-3}}j^{(1)}_i(r)+g^{(1)}_i(r)\\
\end{split}
\end{equation}
Here also at first we determine the constraint equation which is given by
\begin{equation}
\begin{split}
 \frac{\partial_im^{(0)}}{m^{(0)}}+(D-1)\partial_v\beta^{(0)}_i=0
 \end{split}
\end{equation}
This follows from the conservation of boundary stress tensor.\par 
The dynamical equations in the vector sector are the $ri, vi$ components of the Einstein tensor and $i$ th component of the Maxwell equation. Solving these equations with appropriate boundary conditions (regularity at the future event horizon and appropriate fall off at boundary) we have the following expressions for $j^{(1)}(r)$ and $g^{(1)}(r)$.
 \small
 \begin{equation}
\begin{split}
 &j^{(1)}_i(r)=\frac{r~\partial_v\beta^{(0)}_i}{r^2\left(1-V(r)\right)}+\frac{3(D-3)~r^2~Q^{(0)}~\bigg(\partial_iQ^{(0)}+(D-2)~Q^{(0)}~\partial_v\beta^{(0)}_i\bigg)}{R^{2(D-1)}~r^2\left(1-V(r)\right)}R~F_1(\rho,M)\\
 & g^{(1)}_i(r)=\frac{4}{2(D-3)\sqrt{3}Q^{(0)}}\left(-r^{D-2}\partial_v\beta^{(0)}_i+(D-1)m^{(0)}j^{(1)}_i(r)+\left(\frac{1}{4}\frac{3(D-3)}{2(D-2)}\frac{{Q^{(0)}}^{2}}{r^{D-4}}-m^{(0)}r\right)\frac{dj^{(1)}_i(r)}{dr}\right)\\
 \end{split}
\end{equation}
\normalsize
Now plugging these $g^{(1)}_i(r)$ and $j^{(1)}_i(r)$ in the \eqref{correct_vector} we have the corrected first order metric and gauge field in the vector sector as
\begin{equation}
\begin{split}
& g^{(1)}_{vi}(r)=r~\partial_v\beta^{(0)}_i+\frac{3(D-3)~r^2~Q^{(0)}~\bigg(\partial_iQ^{(0)}+(D-2)~Q^{(0)}~\partial_v\beta^{(0)}_i\bigg)}{R^{2(D-1)}}R~F_1(\rho,M)\\
&A^{(1)}_i(r)=-{2}{\sqrt{3}}\frac{r^D}{R^{2(D-1)}}\bigg(\partial_iQ^{(0)}+(D-2)~Q^{(0)}~\partial_v\beta^{(0)}_i\bigg)F_1^{(1,0)}(\rho,M)\\
  \end{split}
\end{equation}
where $F_1$ is given by
\begin{equation}
\begin{split}
&F_1(\rho,M)=\frac{1}{4(D-2)}\left(1+\frac{1}{4}\frac{3(D-3)}{2(D-2)}\frac{Q_1^2}{\rho^{2(D-2)}}-\frac{M}{\rho^{D-1}}\right)F_3(\rho, M)\\\
\text{where}~~~&F_3(\rho, M)=\int_{\rho}^{\infty}dp\frac{1}{\left(1+\frac{1}{4}\frac{3(D-3)}{2(D-2)}\frac{Q_1^2}{p^{2(D-2)}}-\frac{M}{p^{D-1}}\right)^2}\Big(\frac{1}{p^{2(D-1)}}-\frac{c_1}{p^{2D-3}}\Big)\\
\end{split}
\end{equation}
with
\begin{equation}
 c_1=\frac{D-2}{D-1}\left(1+\frac{2}{M(D-3)}\right)
\end{equation}
Where we have used the following rescaled variables
\begin{equation}
\begin{split}
  &\rho=\frac{r}{R}~~, M=\frac{m}{R^{D-1}}~~,Q_{1}=\frac{Q}{R^{D-2}},~~ \text{and}~Q_1^2=4~\frac{2(D-2)}{3(D-3)}(M-1) \\
\end{split}
\end{equation}
And then the Hawking temperature is given by
\begin{equation}
\begin{split}
T=\frac{(D-1)R}{4\pi}\left(1-\left(\frac{D-3}{D-1}\right)\left(M-1\right)\right)
\end{split}
\end{equation}
where $R$ is the radius of the outer horizon. In terms of these rescaled variables the outer horizon is given by $\rho_{+}\equiv 1$.
\subsection{Tensors at first order}
The tensor component of the metric at first order can be parametrized by $\pi^{(1)}_{ij}(r)$ as
\begin{equation}
\begin{split}
 g^{(1)}_{ij}=r^2\pi^{(1)}_{ij}(r)
\end{split}
\end{equation}
 This unknown parameter $\pi^{(1)}_{ij}$ can be determined by solving the dynamical equation obtained from the $ij$ component of Einstein equation, which is given by
\begin{equation}\label{tensor_dynamical_equation}
 \frac{d}{dr}\left(r^DV(r,m,Q)\frac{d\pi_{ij}^{(1)}(r)}{dr}\right)=-2(D-2)r^{D-3}\sigma_{ij}^{(0)}
\end{equation} 
 where
 \begin{equation}
\begin{split}
 \sigma_{ij}^{(0)}=\frac{\partial_i\beta^{(0)}_j+\partial_j\beta^{(0)}_i}{2}-\frac{\partial_k\beta^{(0)}_k}{D-2}\delta_{ij}
 \end{split}
\end{equation}
Demanding regularity at the future event horizon and appropriate fall off at boundary the solution to the equation \eqref{tensor_dynamical_equation} is given by 
\begin{equation}
\begin{split}
\pi^{(1)}_{ij}(r)=\frac{2}{R}\sigma_{ij}~F_2\left(\rho,M\right)
 \end{split}
\end{equation}
where
\begin{equation}
\begin{split}
&F_2\left(\rho,M\right)=\int_{\rho}^{\infty}\frac{\rho^D\left(\rho^D-\rho^2\right)}{\rho^2\left(\rho^{2D}-M\rho^{D+1}+(M-1)\rho^4\right)}d\rho
\end{split}
\end{equation}
\subsection{The global metric and gauge field at first order} 
We have done our computation about a special point $x^{\mu}=0$ in the boundary. However, our perturbation procedure is ultralocal and we could set any arbitrary velocity $u^{\mu}$ to be $\{1, 0, \cdots\}$ by a boundary coordinate transformation. So we could do our computation about any arbitrary point on the boundary. Hence the results recorded in the previous subsections contain enough information to write down the metric and gauge field in covariant form w.r.t the boundary metric. we have the following covariant form of the metric and gauge field 
\begin{equation}
\begin{split}
ds^2=&g_{AB}dx^{A}dx^{B}\\
=&-2u_{\mu}dx^{\mu}dr-r^2V(r, m, Q)u_{\mu}u_{\nu}dx^{\mu}dx^{\nu}+r^2P_{\mu\nu}dx^{\mu}dx^{\nu}\\
&-2u_{\mu}dx^{\mu}~r\left[u^{\lambda}\partial_{\lambda}u_{\nu}-\frac{\partial_{\lambda}u^{\lambda}}{D-2}u_{\nu}\right]dx^{\nu}+\frac{2r^2}{R}F_2(\rho, M)\sigma_{\mu\nu}dx^{\mu}dx^{\nu}\\
&-2u_{\mu}dx^{\mu}\left[\frac{3~(D-3)~Q~r^{2}}{R^{2D-3}}{\cal P}^{\lambda}_{\nu}\left(\mathcal{D}_{\lambda}Q\right)F_1(\rho,M)\right]dx^{\nu}+\cdots\\\\
&A=\left[\frac{\sqrt{3}Q}{2~r^{D-3}}u_{\mu}-\frac{2\sqrt{3}r^{D}}{R^{2(D-1)}}{\cal P}^{\lambda}_{\mu}\left(\mathcal{D}_{\lambda}Q\right)F_1^{(1,0)}(\rho,M)\right]dx^{\mu}+\cdots\\
\end{split}
\end{equation}
where 
\begin{equation}
{\cal P}^{\lambda}_{\mu}\left(\mathcal{D}_{\lambda}Q\right)={\cal P}^{\lambda}_{\mu}\left({\partial}_{\lambda}Q\right)+(D-2)\left(u^{\lambda}\partial_{\lambda}u_{\mu}\right)Q
\end{equation} 
 and
 \begin{equation}
 \sigma_{\mu\nu}={\cal P}_\mu^\alpha {\cal P}_\nu^\beta\left[\frac{\partial_\alpha u_\beta+\partial_\beta u_\alpha}{2}-{\eta}_{\alpha\beta}\left(\frac{\Theta}{D-2}\right)\right]
 \end{equation} 
 And the constraint equations can be written in covariant form as
 \begin{equation}
 \begin{split}
 & \frac{\left(u\cdot\partial\right)Q}{Q}+\partial\cdot u=0\\
  & \frac{\left(u\cdot\partial\right)m}{m}+(D-1)\frac{\partial\cdot u}{D-2}=0\\
  &\frac{{ {\cal P}^{\alpha}_{\mu}~\partial_{\alpha}}m}{m}+(D-1)u^{\lambda}\partial_{\lambda}u_{\mu}=0\\
 \end{split}
\end{equation}
\subsection{The boundary stress tensor and the charge current} 
In this section we write down the expressions for the stress tensor and charge current dual to the metric and gauge field up to first order in derivative expansion. By using AdS/CFT correspondence we can determine the boundary stress tensor from the bulk space-time metric. Here we have to add suitable counter terms to the action to regularise the divergence arising from integrating the full space-time volume. The expression for the boundary stress tensor dual to the metric presented in the previous subsection can be obtained by the prescription of \cite{Balasubramanian1999}.\
\par 
To calculate this stress tensor we need to know the asymptotic expansion of the function $F_2(\rho,M)$, present in the tensor sector of the metric up to $\mathcal{O}\left(\frac{1}{r^D}\right)$. This expansion is given by
\begin{equation}
 \begin{split}
 F_2(r,M)=\frac{R}{r}-\frac{R^{D-1}}{(D-1)r^{D-1}}-\frac{M R^D}{D~r^D}+\cdots
\end{split}
\end{equation}
The stress tensor for the metric up to first order in derivative expansion is given by
\begin{equation}
 \begin{split}
 &T^{\mu\nu}=p\left(\eta^{\mu\nu}+(D-1)u^{\mu}u^{\nu}\right)-2\eta~\sigma^{\mu\nu}\\
 &\text{where}~~~p=\frac{M~R^{D-1}}{16~\pi~G_D}~~\text{and}~~\eta=\frac{R^{D-2}}{16~\pi~G_D}\\
\end{split}
\end{equation}
 The charge current $J_{\mu}$ can be obtained from the gauge field by using the following expression
 \begin{equation}
 \begin{split}
 J_{\mu}=\lim_{r\to\infty}\frac{r^{D-3}A_{\mu}}{8~\pi~G_D}
 \end{split}
\end{equation}
So the charge current for the gauge field is given by
\begin{equation}
 \begin{split}
 J_{\mu}=n~u_{\mu}-\mathfrak{D}~{\cal P}^{\nu}_{\mu}\mathcal{D}_{\nu}n
 \end{split}
\end{equation}
where the charge density $n$ and diffusion constant $\mathfrak{D}$ are given by
\begin{equation}
 \begin{split}
&n=\frac{\sqrt{3}Q}{16~\pi~G_D}~~\text{and}~~\mathfrak{D}=\frac{(D-3)M+2}{R~M~(D-1)(D-3)}
\end{split}
\end{equation}

\subsection{Hydrodynamic metric and gauge field up to first order in derivative}
The metric dual to hydrodynamics in arbitrary dimension $D$ is written in terms of fluid variable $u$, a charge field and a temperature field $T$ living on the $(D-1)$ dimensional boundary of the space-time. The independent data in first order is written in table (\ref{1st_independent_data}). We will write the hydrodynamic metric and the constraint equations in the following way
\begin{table}[ht]
\caption{Data at 1st order in derivative} 
\vspace{0.5cm}
\centering 
\label{1st_independent_data}
\begin{tabular}{|c| c|} 
\hline\hline 
 &Independent Data  \\ [1ex] 
\hline 
\hline
Scalar & $\frac{\Theta}{D-2}\equiv\frac{\partial\cdot u}{D-2}$\\ [1ex]
\hline
Vector &$a_\mu\equiv(u\cdot\partial)u_\mu~,~{\cal P}^{\lambda}_{\mu}\partial_{\lambda}Q_C$\\ [1ex]
\hline
\vspace{-0.3cm}
& \\
Tensor &$\sigma_{\mu\nu}={\cal P}_\mu^\alpha {\cal P}_\nu^\beta\left[\frac{\partial_\alpha u_\beta+\partial_\beta u_\alpha}{2}-{\eta}_{\alpha\beta}\left(\frac{\Theta}{D-2}\right)\right]$\\ [1ex]
\hline
\hline
\end{tabular}
\label{table:1storder} 
\end{table}
\par 

The constraint equations can be written as
\begin{equation}\label{Hconstraint}
 \begin{split}
 &\frac{\left(u\cdot\partial\right)Q_C}{Q_C}=0\\
 & \frac{\left(u\cdot\partial\right)r_H}{r_H}+\frac{\Theta}{D-2}=0\\
 &\frac{{ {\cal P}^{\alpha}_{\mu}~\partial_{\alpha}}r_H}{r_H}+a_{\mu}+f(Q_C){\cal P}^{\alpha}_{\mu}~\partial_{\alpha}Q_C=0\\
 \end{split}
\end{equation}
where
\begin{equation}
f(Q_C)=\frac{6\frac{(D-3)}{D-1}Q_C}{8(D-2)+3(D-3)Q_C^2}
\end{equation}
We write metric and gauge field in two part as
\begin{equation}\label{eq:met}
dS^2 = dS_0^2 + dS_1^2 
\end{equation}
where\\
\textbf{$0^{\text{th}}$ Order Piece:}
\begin{equation}\label{eq:met0}
\begin{split}
dS_0^2 =& -2 u_\mu ~dx^\mu~ dr - r^2 V(r)~u_\mu u_\nu~dx^\mu ~dx^\nu+r^2 {\cal P}_{\mu\nu} dx^\mu dx^\nu\\
&{\cal P}_{\mu\nu}= \eta_{\mu\nu}+u_\mu u_\nu
\end{split}
\end{equation}
\textbf{$1^{\text{st}}$ Order Piece:}
\begin{equation}\label{eq:met1}
\begin{split}
&dS_1^2 = -r \left(u_\mu B_\nu + u_\nu B_\mu\right) dx^\mu~dx^\nu + \frac{2 r^2}{r_H}~ F_2(\rho,M)~\sigma_{\mu\nu} ~dx^\mu~dx^\nu\\
&\text{where}\\
&B_\mu =a_\mu -\left(\Theta\over D-2\right) u_\mu+\frac{3~(D-3)~r~Q_C}{r_H^2}{\cal P}^{\lambda}_{\mu}\left({\partial}_{\lambda}Q_C\right)\bigg[1-(D-2)Q_C~f(Q_C)\bigg]F_1(\rho,M)\\
\end{split}
\end{equation}
where 
\begin{equation}
{\cal P}^{\lambda}_{\mu}\left(\mathcal{D}_{\lambda}Q\right)=r_H^{(D-2)}\bigg[1-(D-2)Q_C~f(Q_C)\bigg]~{\cal P}^{\lambda}_{\mu}\left({\partial}_{\lambda}Q_C\right)\\
\end{equation} 
The gauge field up to first order in derivative expansion can be written as 
 \begin{equation}\label{eq:gauge_field}
A = A_0 + A_1 
\end{equation}
where\\
\textbf{$0^{\text{th}}$ Order Piece:}
\begin{equation}\label{eq:gauge_field0}
\begin{split}
& A_0 =\frac{\sqrt{3}~r~Q_C}{2}\left(\frac{r_H}{r}\right)^{D-2}u_{\mu}dx^{\mu}\\
\end{split}
\end{equation}
\textbf{$1^{\text{st}}$ Order Piece:}
\begin{equation}\label{eq:gauge_field1}
\begin{split}
& A_1 = -2\sqrt{3}\left(\frac{r}{r_H}\right)^D\bigg[1-(D-2)Q_C~f(Q_C)\bigg]F_1^{(1,0)}(\rho,M)~{\cal P}^{\lambda}_{\mu}\left({\partial}_{\lambda}Q_C\right)dx^{\mu}\\
\end{split}
\end{equation}
where, 
\begin{equation}
V(r)=1-\left(1+\frac{1}{4}~\frac{3(D-3)}{2(D-2)}Q_C^2\right)\left(\frac{r_H}{r}\right)^{D-1}+\frac{1}{4}~\frac{3(D-3)}{2(D-2)}Q_C^2\left(\frac{r_H}{r}\right)^{2(D-2)}
\end{equation}
And $r=r_H$ is the position of the outer event horizon of the space-time.
\section{The large $D$ metric, gauge field and membrane equations:}\label{largeDmetricgauge}
In this section we simply quote the results of large $D$ metric and gauge field as in paper \cite{Kundu2018chargeAdsds}.\\
The metric in the large $D$ expansion is written in a split form as $\bar W_{AB}$ plus $W^{(rest)}_{AB}$.
$$W_{AB} = \bar W_{AB} + W_{AB}^{(rest)}$$
 $\bar W_{AB}$ is the metric of the nonsingular asymptotic geometry. In our case it would be the metric of just pure AdS. In general, $\bar W_{AB}$ is any smooth  solution (where all components of the Riemann tensor are of order ${\cal O}(1)$ in terms of large $D$ counting) to Einstein's equations with cosmological constant and vanishing electromagnetic field.
$W^{(rest)}_{AB}$ captures the effect of black hole and singularity.
Using the large $D$ technique one can determine $W_{AB}^{(rest)}$ and in this case also the gauge field $A_M$ order by order in an expansion in inverse powers of $D$. In other words the metric an the gauge field will have the following form
\begin{equation}
\begin{split}
 &W^{(rest)}_{AB}=\mathcal{W}^{(0)}_{AB}+\left(\frac{1}{D}\right)\mathcal{W}^{(1)}_{AB}+\cdots\\
 &A_M=A^{(0)}_M+\left(\frac{1}{D}\right)A^{(1)}_M+\cdots
\end{split}
\end{equation} 
\\
The final solution is expressed in terms of a null geodesic field $O^A\partial_A$ and two scalar fields $\psi$ and $\tilde Q$.
$\psi$ is a smooth function with $\psi=1$ as event horizon of the full space-time and harmonic w.r.t the background $\bar{W}_{AB}$. $\widetilde{Q}$ is also a smooth function satisfying
\begin{equation}\label{subQ}
\begin{split}
n\cdot\partial\widetilde{Q}=0~,~\text{and}~\widetilde{Q}|_{\psi=1}=\frac{1}{\sqrt{2}}\left(U^MA_M\right)|_{\psi=1}
  \end{split}
\end{equation}
where $n_A$ is the unit normal to the constant $\psi$ slices and $U$ is the membrane velocity defined by $U=n-O$.\\
The gauge is fixed by demanding that  to all orders
 $$O^A W^{(rest)}_{AB}=0 ~~\text{and}~~ O^MA_M =0$$
 
Now we shall present the final solutions upto the first subleading order in ${\cal O}\left(1\over D\right)$ i.e., the explicit expressions for $\mathcal{W}^{(0)}_{AB}~$, $A^{(0)}_M$ and $\mathcal{W}^{(1)}_{AB}~$, $A^{(1)}_M$ 
\begin{equation}
\begin{split}
 &\mathcal{W}^{(0)}_{AB}=fO_AO_B\\
 &A^{(0)}_M=\sqrt{2}~\widetilde{f}O_M\\
 &\mathcal{W}^{(1)}_{AB}=\mathcal{Z}^{(s1)}O_AO_B+\left(\mathcal{Z}^{(v)}_AO_B+\mathcal{Z}^{(v)}_BO_A\right)+\mathcal{Z}^{(T)}_{AB}\\
 \text{and}~&A^{(1)}_M=\mathcal{A}^{(s)}O_M+\mathcal{A}^{(v)}_M
\end{split}
\end{equation} 
where
 \begin{equation}
\begin{split}
 f=\left(1+\widetilde{Q}^2\right)\psi^{-D}~,~~\widetilde{f}=\widetilde{Q}\psi^{-D}
 \end{split}
\end{equation} 
 and
 \begin{equation}
\begin{split}
 &\mathcal{Z}^{(s1)}=\sum_{i=1}^{N_S}S^{(i)}_1(\zeta)\mathcal{S}^{(i)}~,~\mathcal{Z}^{(s2)}=\sum_{i=1}^{N_S}S^{(i)}_2(\zeta)\mathcal{S}^{(i)}~,~\mathcal{A}^{(s)}=\sum_{i=1}^{N_S}a^{(i)}_s(\zeta)\mathcal{S}^{(i)}\\
 &\mathcal{Z}^{(v)}_A=\sum_{i=1}^{N_V}\mathcal{V}^{(i)}(\zeta){V}^{(i)}_A~,~\mathcal{A}^{(v)}_A=\sum_{i=1}^{N_V}a^{(i)}_v(\zeta){V}^{(i)}_A~,~\mathcal{Z}^{(T)}_{AB}=\sum_{i=1}^{N_T}\mathcal{T}^{(i)}(\zeta){t}^{(i)}_{AB}
 \end{split}
\end{equation}  
Here $O_A~dX^{A}$ is a null  one-form with respect to both the background metric $\bar{W}_{AB}$ and full metric $W_{AB}$. This is dual to the geodesic vector field $O^A\partial_A$ mentioned before. The function $\psi$ and $\tilde Q$ are already defined (see the discussion around equation \eqref{subQ}).\\
\par
Now both the functions $\psi$ and $\widetilde Q$ admit ${1\over D}$ expansion in the `membrane region' - the region where the gravitational field is non trivial in the limit of large $D$. It follows that upto first order in $\frac{1}{D}$, the function $f$ and $\widetilde{f}$ can be written as
\begin{equation}
\begin{split}
&f(W)=\left(1+\widetilde{Q}^2\right)~e^{-W}-\widetilde{Q}^2~e^{-2W}+\left(\frac{1}{D}\right)\\
&\widetilde{f}(W)=\widetilde{Q}~e^{-W}+\left(\frac{1}{D}\right)
\end{split}
\end{equation}
where the parameter $W$ is a $\mathcal{O}(1)$ variable, defined by $W=D\left(\psi-1\right)$.\par
 And then the solution up to first order in metric and gauge field can be written as
\begin{equation}
\begin{split}
&\mathcal{V}^{(i)}(y)=-2\int_{y}^{\infty}dx~e^{-x}~\widetilde{Q}~a^{(i)}_{v}(x)-e^{-y}\mathcal{K}_{vector}-2\left(\frac{D}{K}\right)\int_{y}^{\infty}dx~e^{-x}\int_{0}^{x}dt\frac{e^{t}~\mathfrak{v}^{(i)}_{\text{metric}}(t)}{1-f(t)}\\
&\text{where}~\mathcal{K}_{vector}=-2\int_{0}^{\infty}dx~e^{-x}~\widetilde{Q}~a^{(i)}_{v}(x)-2\left(\frac{D}{K}\right)\int_{0}^{\infty}dx~e^{-x}\int_{0}^{x}dt\frac{e^{t}~\mathfrak{v}^{(i)}_{\text{metric}}(t)}{1-f(t)}
\end{split}
\end{equation}
and
\begin{equation}
\begin{split}
a^{(i)}_v(t)&=-e^t(f-\widetilde{f}^2)\int_t^{\infty}dx\frac{e^{-3x}}{(1-f)(f-\widetilde{f}^2)^2}\int_0^{x}dy~M^{(i)}(y)\\
&+\left(2\int_0^{\infty}dz~\widetilde{f}~a^{(i)}_v(z)\right)e^t(f-\widetilde{f}^2)\int_t^{\infty}dx\frac{e^{-3x}}{(1-f)(f-\widetilde{f}^2)^2}\int_0^{x}dy~e^y~\widetilde{f}~(f-\widetilde{f}^2)\\
\end{split}
\end{equation}
where
\begin{equation}
\begin{split}
M^{(i)}(x)=&\int_0^xdx~e^{2x}\left(f-\widetilde{f}^2\right)\Bigg(\widetilde{f}~e^{-x}\left(-\frac{2D}{K}\right)\int_0^{\infty}dz~e^{-z}\int_0^{z}dt\frac{e^{t}~\mathfrak{v}^{(i)}_{\text{metric}}(t)}{1-f(t)}-\frac{1}{N}\mathfrak{v}^{(i)}_{\text{gauge}}(x)\\
&+2~\widetilde{f}\left(\frac{D}{K}\right)e^{-x}\int_0^xdt~\frac{e^{t}~\mathfrak{v}^{(i)}_{\text{metric}}(t)}{1-f(t)}\Bigg)
\end{split}
\end{equation}
and
\begin{equation}
\begin{split}
&\int_0^{\infty}dt~\widetilde{f}~a^{(i)}_v=\frac{A}{B}\\
\text{where}~&B=1-2\int_0^{\infty}dz~\widetilde{Q}~(f-\widetilde{f}^2)\int_z^{\infty}dx\frac{e^{-3x}}{(1-f)(f-\widetilde{f}^2)^2}\int_0^{x}dy~e^y~\widetilde{f}~(f-\widetilde{f}^2)\\
\text{And}~&A=-\int_0^{\infty}dz~\widetilde{Q}~(f-\widetilde{f}^2)\int_z^{\infty}dx\frac{e^{-3x}}{(1-f)(f-\widetilde{f}^2)^2}\int_0^{x}dy~M^{(i)}(y)
\end{split}
\end{equation}
and
\begin{equation}
\begin{split}
a^{(i)}_s(y)=-e^{-y}\left(\frac{1}{N}\right)\int_0^{\infty}~d\rho~e^{-\rho}\int_0^{\rho}d\zeta~e^{\zeta}~s^{(i)}_{\text{gauge}}(\zeta)+\left(\frac{1}{N}\right)\int_y^{\infty}d\rho~e^{-\rho}\int_0^{\rho}d\zeta~e^{\zeta}~s^{(i)}_{\text{gauge}}(\zeta)
\end{split}
\end{equation}
and
\begin{equation}
\begin{split}
&S^{(i)}_1(y)=-4\int_y^{\infty}d\rho~\widetilde{f}~a^{(i)}_s(\rho)-e^{-y}~A_{scalar}+\left(\frac{2}{N}\right)\int_y^{\infty}d\rho~e^{-\rho}\int_0^{\rho}d\zeta~e^{\zeta}~s^{(i)}_{\text{metric}}(\zeta)\\
\text{where}~&A_{scalar}=-4\int_0^{\infty}d\rho~\widetilde{f}~a^{(i)}_s(\rho)+\left(\frac{2}{N}\right)\int_0^{\infty}d\rho~e^{-\rho}\int_0^{\rho}d\zeta~e^{\zeta}~s^{(i)}_{\text{metric}}(\zeta)\\
\end{split}
\end{equation}
and correction to the tensor sector
\begin{equation}
\begin{split}
{\cal T}\left(W\right)=2\frac{D}{K}\text{log}\left[1-\widetilde{Q}^2e^{-W}\right]
\end{split}
\end{equation}
where
\begin{equation}
\begin{split}
&\mathfrak{v}^{(1)}_{\text{metric}}=\frac{\widetilde{f}^2-f}{2}~,~\mathfrak{v}^{(2)}_{\text{metric}}=\mathfrak{v}^{(4)}_{\text{metric}}=\frac{f}{2}~,~\mathfrak{v}^{(~3)}_{\text{metric}}=-\frac{\widetilde{f}^2}{2}\\
&\mathfrak{v}^{(1)}_{\text{gauge}}=\widetilde{f}~,~\mathfrak{v}^{(2)}_{~\text{gauge}}=-\widetilde{f}~,~\mathfrak{v}^{(~4)}_{\text{gauge}}=-\widetilde{f}
\end{split}
\end{equation}
and
\begin{equation}
\begin{split}
&{s}^{(1)}_{\text{metric}}=\frac{N}{2}\left(\dot{\cal V}^{(2)}+\dot{\cal V}^{(3)}+\dot{\cal V}^{(4)}\right)-\widetilde{f}^2~,~{s}^{(2)}_{\text{metric}}=\frac{N}{2}\left(-\dot{\cal V}^{(2)}+\dot{\cal V}^{(3)}\right)+\widetilde{f}^2\\
&{s}^{(3)}_{\text{metric}}=\frac{f-\widetilde{f}^2}{2}~,~{s}^{(4)}_{\text{metric}}=\frac{N}{2}\dot{\cal V}^{(1)}~,~{s}^{(5)}_{\text{metric}}=\left(\widetilde{Q}\widetilde{f}-\widetilde{f}^2\right)~,~{s}^{(6)}_{\text{metric}}=0\\
&{s}^{(7)}_{\text{metric}}=-\widetilde{f}^2~,~{s}^{(8)}_{\text{metric}}=\frac{N}{2}\left(-\dot{\cal V}^{(2)}+\dot{\cal V}^{(3)}\right)\\
&s^{(1)}_{\text{gauge}}=N\left(\widetilde{f}{\cal V}^{(2)}+f\dot{a}^{(2)}_v+\widetilde{f}{\cal V}^{(4)}+f\dot{a}^{(4)}_v\right)~,~s^{(2)}_{\text{gauge}}=N\left(-\widetilde{f}{\cal V}^{(2)}-f\dot{a}^{(2)}_v+\widetilde{f}{\cal V}^{(3)}+f\dot{a}^{(3)}_v\right)\\
&s^{(4)}_{\text{gauge}}=N\left(\widetilde{f}{\cal V}^{(1)}+f\dot{a}^{(1)}_v\right)~,~s^{(6)}_{\text{gauge}}=\widetilde{f}~,~s^{(3)}_{\text{gauge}}=s^{(5)}_{\text{gauge}}=s^{(7)}_{\text{gauge}}=0\\
&s^{(8)}_{\text{gauge}}=N\left(-\widetilde{f}{\cal V}^{(2)}-f\dot{a}^{(2)}_v+\widetilde{f}{\cal V}^{(3)}+f\dot{a}^{(3)}_v\right)+\widetilde{f}
\end{split}
\end{equation}
\begin{table}[ht!]
\caption{Membrane Data} 
\vspace{0.5cm}
\centering 
\begin{tabular}{|c| c| c|} 
\hline\hline 
 Scalar&Vector&Tensor  \\ [1ex] 
\hline 
\hline
\vspace{-0.3cm}
&  &  \\
${\cal S}^{(1)}\equiv\frac{\left(U\cdot\nabla\right) K}{K}$&${ V}^{(1)}_{A}\equiv P^C_A\left(\frac{\nabla_C K}{K}\right)$& $t_{AB}\equiv P^C_A P^{C^{\prime}}_B\left[\left(\frac{\nabla_C O_{C^{\prime}}+\nabla_{C^{\prime}} O_{C}}{2}\right)-\frac{P_{CC^{\prime}}}{D}\left(\nabla\cdot O\right)\right]$  \\ [1ex]
\hline
\vspace{-0.3cm}
& & \\
${\cal S}^{(2)}\equiv U\cdot K\cdot U $&${ V}^{(2)}_{A}\equiv  P^C_A\left(U\cdot\nabla\right)O_C $& \\ [1ex]
\hline
\vspace{-0.3cm}
& & \\
${\cal S}^{(3)}\equiv\hat{\nabla}\cdot U$& ${V}^{(3)}_{A}\equiv P^C_A\left(U\cdot\nabla\right)U_C$&\\ [1ex]
\hline
\vspace{-0.3cm}
& &\\
${\cal S}^{(4)}\equiv \frac{\hat{\nabla}^2K}{K^2}$&${V}^{(4)}_{A}\equiv P^C_A\left(\frac{\hat{\nabla}^2U_C}{K}\right)$&\\ [1ex]
\hline
\vspace{-0.3cm}
&& \\
${\cal S}^{(5)}\equiv U\cdot\left(\frac{\nabla\widetilde{Q}}{\widetilde{Q}}\right)$&&\\ [1ex]
\hline
\vspace{-0.3cm}
&& \\
${\cal S}^{(6)}\equiv\frac{1}{K} \nabla\cdot\left(\frac{\nabla\widetilde{Q}}{\widetilde{Q}}\right)$&&\\ [1ex]
\hline
\vspace{-0.3cm}
&& \\
${\cal S}^{(7)}\equiv \frac{K}{D}$&&\\ [1ex]
\hline
\vspace{-0.3cm}
&& \\
${\cal S}^{(8)}\equiv \frac{R_{uu}}{K}$&&\\ [1ex]
\hline
\hline
\end{tabular}
\label{table:scalar_larged} 
\end{table} 
\subsection{The dual system}
The large-$D$ gravity solutions, described in the  previous subsection, are dual to a co-dimension one, massive and charged, membrane embedded in the asymptotic geometry (AdS for our purpose).
The membrane is characterized by a velocity field $U$, named as `membrane velocity', a charge field $\widetilde{Q}$ and a shape function $\psi$ (the same scalar fields that appear in the bulk metric and gauge fields). The function $\psi$ is a harmonic function with respect to the background geometry. Just like in fluid gravity correspondence, here also   the velocity field $U$ and the shape function $\psi$ cannot be chosen arbitrarily. They have to satisfy some constraint equations, which we shall refer to as membrane equations. For every solution to these membrane equations, we have one solution to Einstein's equations.
The membrane equations are given by
\begin{equation}\label{memceq}
\begin{split}
&P^{A}_C\left[\frac{\widehat{\nabla}^2U_A}{K}-\left(1+\widetilde{Q}^2\right)\frac{\widehat{\nabla}_AK}{K}+U^{B}K_{BA}-\left(1+\widetilde{Q}^2\right)\left(U\cdot\widehat{\nabla}U_A\right)\right]=\mathcal{O}\left(\frac{1}{D}\right)\\
&\hat{\nabla}\cdot U=\mathcal{O}\left(\frac{1}{D}\right)\\
&\frac{\widehat{\nabla}^2\widetilde{Q}}{K}-U\cdot\widehat{\nabla}\widetilde{Q}-\widetilde{Q}\left[\frac{U\cdot\widehat{\nabla}K}{K}-U\cdot K\cdot U-\frac{R_{uu}}{K}\right]=\mathcal{O}\left(\frac{1}{D}\right)
\end{split}
\end{equation}
where
\begin{equation}
\begin{split}
P_{AB}=\bar{W}_{AB}-n_An_B+U_AU_B~,~{R}_{uu}=U^A\bar{R}_{AB}U^B~,~\text{and}~\bar{R}_{AB}=\left(D-1\right)\lambda \bar{W}_{AB}
\end{split}
\end{equation}
\section{Comparing fluid-gravity and membrane-gravity dualities :}\label{sec:gr_solution_comp}
In this section we compare the two perturbation techniques, `derivative expansion' and `large-$D$ expansion' which are used to generate dynamical black-brane solutions to Einstein equations.\\
 `Derivative expansion' is used to solve Einstein's equations with negative cosmological constant, whereas large-$D$ expansion technique is used to solve Einstein equations with or without cosmological constant. `Derivative expansion' generate gravity solutions that are dual to the relativistic Navier-Stokes equations of fluid dynamics. On the other hand large-$D$ expansion techniques generate solutions that are dual to a co-dimension one dynamical membrane embedded in some background space. Like in the previous papers \cite{Bhattacharyya2019leading,Bhattacharyya2019sunleading}, here also our goal is to compare these two gravity solutions along with their dual systems for the charged case. We will show that in appropriate regime of parameter space there exists an overlap regime between these two different looking gravity solutions generated by two different perturbation techniques, which we can see after a coordinate transformation.\par  
\subsection{The split of the hydrodynamic metric}\label{sec:field_redefinition}
As we have mentioned earlier, the metric generated in large-$D$ expansion technique are written in a split form, background plus `rest'. Here we have a null geodesic, which when contracted with the `rest' part vanishes to all order in $\frac{1}{D}$. This is not the case for hydrodynamic metric.\\
  So to compare these two solutions, the first step would be  to split the hydrodynamic metric into background plus `rest'. 
  
  We shall do it in the following way.\\
  
First we find out an affinely parametrized null geodesic $\bar{O}^A\partial_A$ w.r.t the full space-time metric which passes thorough the event horizon of the space-time. This null geodesic vector is normalized in a way such that $O^An_A=1$ ($O_A$ is related to $\bar{O}_A$ by an overall normalization constant), everywhere in the background, where $n_A$ is the unit normal to the constant $\psi$ hypersurfaces. Now the hydrodynamic metric is written in a gauge where $\mathcal{G}_{rr}=0, ~\text{and}~\mathcal{G}_{r\mu}=-u_{\mu}$ to all order in derivative expansion. In this gauge $k^A\partial_A=\zeta(x)\partial_r$ is an affinely parametrized null geodesic to all order in derivative expansion. However we have to normalize this null geodesic and hence we can set $\zeta(x)$ to be one. Ultimately we will find that $\bar{O}^A\partial_A=\partial_r$ is the null geodesic which split the hydrodynamic metric as we want. After that we will choose a coordinate system $Y^A\equiv\left(\rho,y^{\mu}\right)$ where the background of hydrodynamic metric take the following form
\begin{equation}\label{pure_ads}
\begin{split}
dS^2_{\text{background}}=\bar{G}_{AB}dY^AdY^B=\frac{d\rho^2}{\rho^2}+\rho^2\eta_{\mu\nu}
\end{split}
\end{equation}
The $Y^A$ coordinates are related to $X^A\equiv\left(r,x^{\mu}\right)$ coordinate by the mapping $f$
\begin{equation}
\begin{split}
Y^A=f^A\left(X\right)
\end{split}
\end{equation}
We can determine this mapping function by the following equation
\begin{equation}\label{map_eq}
\begin{split}
\bar{O}^A\mathcal{G}_{AB}|_{\{X\}}=\bar{O}^A\left(\frac{\partial f^C}{\partial X^A}\right)\left(\frac{\partial f^{{C'}}}{\partial X^{B}}\right)\bar{G}_{CC'}|_{\{X\}}
\end{split}
\end{equation}
If we use the fact that $\bar{O}^A\partial_A=\partial_r$, then \eqref{map_eq} can be written as 
\begin{equation}
\begin{split}
\mathcal{G}^{(rest)}_{rB}=0
\end{split}
\end{equation}
where $\mathcal{G}^{(rest)}_{AB}=\left(\mathcal{G}_{AB}-\mathcal{\bar{G}}_{AB}\right)$, all written in $\{X^A\}$ coordinates.
As previously noted the hydrodynamic metric metric is written in a particular gauge $\mathcal{G}_{rr}=0, \text{and}~\mathcal{G}_{r\mu}=-u_{\mu}$ to all order in derivative expansion. We will find that coordinate transformation of the form
\begin{equation}
\begin{split}
\rho=r+\chi(x)~~~\text{and}~~~y^{\mu}=x^{\mu}+\frac{u^{\mu}}{r+\chi(x)}+\zeta^{\mu}(x)
\end{split}
\end{equation}
where $u_{\mu}\partial_{\nu}\zeta^{\mu}=0$, keep the hydrodynamic metric in this required gauge. Further it will turn out that for the exact matching of the two metric and gauge field we should have $\zeta^{\mu}=0$ and $\chi(x)=-\frac{\Theta}{D-2}$. So finally we have\footnote{For a detail discussion see \cite{Bhattacharyya2019sunleading}.}
\begin{equation}
\begin{split}
\rho=r-\frac{\Theta}{D-2}~~~\text{and}~~~y^{\mu}=x^{\mu}+\frac{u^{\mu}}{r-\frac{\Theta}{D-2}}
\end{split}
\end{equation}
If we apply these coordinate transformations, the background metric in $\{X^A\}$ coordinate can be written as\footnote{The inverse of the background metric and the christoffel symbols w.r.t background metric are give in Appendix \ref{bacground_christoffel}}
\begin{equation}
\begin{split}
\mathcal{\bar{G}}_{rr}&=0\\
\mathcal{\bar{G}}_{\mu r}&=-u_{\mu}\\
\mathcal{\bar{G}}_{\mu\nu}&=r^2\left({\cal P}_{\mu\nu}-u_{\mu}u_{\nu}\right)-r\left(u_{\mu}a_{\nu}+u_{\nu}a_{\mu}\right)+2r\sigma_{\mu\nu}+2r\frac{\Theta}{D-2}u_{\mu}u_{\nu}+\mathcal{O}\left(\partial^2\right)\\
\end{split}
\end{equation}
Once we know the background metric in $\{X^{A}\}$ coordinates, by subtracting it from the full metric we can determine $\mathcal{G}^{(rest)}_{AB}$. Now by our construction $\mathcal{G}^{(rest)}_{rr}$ and $\mathcal{G}^{(rest)}_{r\mu}$ are identically zero to all order in derivative expansion and the $\mathcal{G}^{(rest)}_{\mu\nu}$ component can be written as
\begin{equation}
\begin{split}
\mathcal{G}^{\left(rest\right)}_{\mu\nu}&=\mathcal{G}^{(S1)}u_{\mu}u_{\nu}+\mathcal{G}^{(S2)}{\cal P}_{\mu\nu}+\left(\mathcal{G}^{(V)}_{\mu}~u_{\nu}+\mathcal{G}^{(V)}_{\nu}~u_{\mu}\right)+\mathcal{G}^{(T)}_{\mu\nu}\\
\text{where,}\\
&\mathcal{G}^{(S1)}=r^2\bigg(1-V(r)\bigg)\\
&\mathcal{G}^{(S2)}=\mathcal{O}\left(\partial^2\right)\\
&\mathcal{G}^{(V)}_{\mu}=-\frac{3(D-3)~r^2~Q_C}{r_H^2}\bigg[1-(D-2)Q_Cf(Q_C)\bigg]F_1(\rho,M)~{\cal P}^{\lambda}_{\mu}\left({\partial}_{\lambda}Q_C\right)\\
&\mathcal{G}^{(T)}_{\mu\nu}=2r\left(\frac{r}{r_H}F_2\left(\rho,M\right)-1\right)\sigma_{\mu\nu}
\end{split}
\end{equation}
\subsection{Membrane data in terms of fluid data}
In derivative expansion the solutions are characterized by a velocity filed $u$, called fluid velocity, a temperature field $T$ and  a charge field $Q$, whereas in large-$D$ expansion characterising data are the shape function $\psi$, the charge $\tilde Q$ and the membrane velocity $U$. The number of data does match on both sides, as it should be. But these variables are not the same and we need to rewrite one in terms of the other, to perform a comparison. In this subsection we rewrite the characterising data of the membrane in terms of the fluid variables.\\
 \subsubsection{Determining $\psi$}
As we have described before,
  $\psi$ is a  scalar function, harmonic with respect to  the background geometry. The hypersurface $\psi=1$ is identified with the dynamical horizon of  the black brane solution. So we have to solve the differential equation $\nabla^2\psi^{-D}=0$ in this background geometry order by order order in both the perturbation parameters.\par
  
   After solving we find the following expression for the function $\psi$ (see appendix-B of \citep{Bhattacharyya2019leading} for the details of the calculation)
\begin{equation}\label{psi_redefinition}
\psi\left(r,x^{\mu}\right)=1+\left(1-\frac{1}{D}\right)\left(\frac{r}{r_H}-1\right)+\mathcal{O}\left(\partial^2,\frac{1}{D^3}\right)
\end{equation}
\subsubsection{Determining $U^A$}
Once we have the $\psi$ field everywhere, we could compute the unit normal to the constant $\psi$ surfaces.
\begin{equation}
\begin{split}
&n_r=\frac{1}{r}+\mathcal{O}\left(\partial^2\right)~~~~~~~~~~~~~~~~~~~n_{\mu}=-\frac{\Theta}{D-2}u_{\mu}+a_{\mu}+f(Q_C){\cal{P}}^{\alpha}_{\mu}\partial_{\alpha}Q_C+\mathcal{O}\left(\partial^2\right)\\
&n^r=r-\frac{\Theta}{D-2}+\mathcal{O}\left(\partial^2\right)~~~~~~~n^{\mu}=\frac{u^{\mu}}{r}+\frac{f(Q_C)}{r^2}~{\cal P}^{\mu\alpha}~\partial_{\alpha} Q_C+\mathcal{O}\left(\partial^2\right)\\
\end{split}
\end{equation}
Now $U^A$ is defined as follows
\begin{equation}
U_{A}=n_A-O_A
\end{equation}
After properly normalizing our null geodesic field by $O^A=\frac{\bar{O}^A}{\left(n_A\cdot\bar{O}^A\right)}$~, we have
\begin{equation}\label{O_redefinition}
\begin{split}
&O_r=0~~~~~~~~~~~~~~~~~~~~~~O_{\mu}=-ru_{\mu}+\mathcal{O}\left(\partial^2\right)\\
&O^r=r+\mathcal{O}\left(\partial^2\right)~~~~~~~~~O^{\mu}=0\\
\end{split}
\end{equation}
Then we can find out the membrane velocity $U_A$ as
\begin{equation}\label{U_redefinition}
\begin{split}
&U_r=\frac{1}{r}~~~~~~~~~~~~~~~~~~~~~~~~~~~~~U_{\mu}=ru_{\mu}-\frac{\Theta}{D-2}u_{\mu}+a_{\mu}+f(Q_C){\cal{P}}^{\alpha}_{\mu}\partial_{\alpha}Q_C+\mathcal{O}\left(\partial^2\right)\\
&U^r=-\frac{\Theta}{D-2}+\mathcal{O}\left(\partial^2\right)~~~~~~~U^{\mu}=\frac{u^{\mu}}{r}+\frac{f(Q_C)}{r^2}~{\cal P}^{\mu\alpha}~\partial_{\alpha}Q_C+\mathcal{O}\left(\partial^2\right)\\
\end{split}
\end{equation}

\subsubsection{Determining $\widetilde{Q}$}
Next our goal is to write the smooth function $\widetilde{Q}$ present in the large $D$ metric and gauge field in terms of fluid data. This function satisfies the subsidiary condition $\left(n\cdot\nabla\right)\widetilde{Q}=0$. The boundary condition which fix it completely is given by
\begin{equation}
\begin{split}
\widetilde{Q}|_{\psi=1}&=\frac{1}{\sqrt{2}}\left(U^MA_{M}\right)|_{\psi=1}\\
&=-\frac{1}{\sqrt{2}}~\frac{\sqrt{3}Q}{2~r^{D-2}}|_{\psi=1}\\
&=-\frac{\sqrt{3}Q}{2\sqrt{2}~r_H^{D-2}}\\
\end{split}
\end{equation}
We want to solve $\widetilde{Q}$ such that $\left(n\cdot\partial\right)\widetilde{Q}=0$ and for that we will take the following expansion in $\widetilde{Q}$.
\begin{equation}
\begin{split}
&\widetilde{Q}=\widetilde{Q}_0+\widetilde{Q}_1~\left(r-r_H\right)+\cdots\\
\text{where}~~~&\widetilde{Q}_0=-\frac{\sqrt{3}Q}{2\sqrt{2}~r_H^{D-2}}\\
\end{split}
\end{equation}
 Collecting coefficients of $\left(r-r_H\right)^0$ after applying $\left(n\cdot\partial\right)\widetilde{Q}=0$ and using the fact that $\left(n\cdot\partial\right)\widetilde{Q}_0=\frac{\sqrt{3}Q}{2\sqrt{2}~r~r_H^{D-2}}\left(\frac{\left(u\cdot\partial\right)Q}{Q}+\Theta\right)=0$, we have $\widetilde{Q}_1=0$. 
So we have $$\widetilde{Q}=-\frac{\sqrt{3}Q}{2\sqrt{2}~r_H^{D-2}}+\mathcal{O}\left(\partial^2\right)$$
and finally we have
\begin{equation}
\widetilde{Q}=-\frac{\sqrt{3}~Q_C}{2\sqrt{2}}+\mathcal{O}\left(\partial^2\right)
\end{equation}

\subsubsection{Relevant derivatives of the basic data}
Large-$D$ metric is determined in terms of the basic functions $\psi$, $\widetilde Q$ and $U^A$ and their derivatives with respect to the induced coordinates on the membrane.  In this subsection we shall convert these `membrane derivatives' of the basic `membrane data' in terms of the fluid data.\par

One of the key structure that arises repeatedly in large-$D$ construction  is the extrinsic curvature of the membrane, viewed as a hypersurface embedded in the background.
The expressions for extrinsic curvature can be re-expressed in terms of fluid variables as
\begin{equation}
\begin{split}
&K_{rr}=-\frac{1}{r^2}+\mathcal{O}\left(\partial^2\right)\\
&K_{r\mu}=-u_{\mu}+\frac{1}{r}\left(\frac{\Theta}{D-2}u_{\mu}-a_{\mu}\right)+\mathcal{O}\left(\partial^2\right)\\
&K_{\mu\nu}=r^2\left({\cal P}_{\mu\nu}-u_{\mu}u_{\nu}\right)+2~r\left(\frac{\Theta}{D-2}u_{\mu}u_{\nu}-\frac{u_{\mu}a_{\nu}+u_{\nu}a_{\mu}}{2}+\sigma_{\mu\nu}\right)+\mathcal{O}\left(\partial^2\right)\\
&K=\left(D-1\right)+\mathcal{O}\left(\partial^2\right)
\end{split}
\end{equation}
where $K_{AB}$ is defined as
\begin{equation}
\begin{split}
&K_{AB}=\Pi_{A}^C~\nabla_C~n_B\\
\text{with}~~~~&\Pi_{AB}=\bar{\mathcal{G}}_{AB}-n_A~n_B\\
\end{split}
\end{equation}
The rest of the data that are relevant for our purpose are presented in the tables -(\ref{scalar_data}), (\ref{vector_data}) and (\ref{tensor_data}). Here $\hat{\nabla}$ is defined for a general $n$ index tensor $X_{A_1A_2\cdots A_n}$ as
\begin{equation}
\hat{\nabla}_AX_{A_1A_2\cdots A_n}=\Pi^{C}_{A}\Pi^{C_1}_{A_1}\Pi^{C_2}_{A_2}\cdots \Pi^{C_n}_{A_n}{\nabla}_CX_{C_1C_2\cdots C_n}
\end{equation}

\begin{table}[hp]
\caption{Scalar large-$D$ Data in terms of fluid Data} 
\vspace{0.5cm}
\centering 
\label{scalar_data}
\begin{tabular}{|c| c|} 
\hline\hline 
 Large-$D$ Data&Corresponding Fluid Data  \\ [1ex] 
\hline 
\hline
\vspace{-0.3cm}
& \\
${\mathcal{ S}}^{(1)}\equiv\frac{\left(U\cdot\nabla\right) K}{K}$ & $=~0$\\ [1ex]
\hline
\vspace{-0.3cm}
& \\
${\mathcal{S}}^{(2)}\equiv U\cdot K\cdot U $&$=~-1$\\ [1ex]
\hline
\vspace{-0.3cm}
& \\
${\mathcal{S}}^{(3)}\equiv\hat{\nabla}\cdot U=\Pi^{AB}(\nabla_A U_B)$&$=~0$\\ [1ex]
\hline
\vspace{-0.3cm}
& \\
${\mathcal{S}}^{(4)}\equiv \frac{\hat{\nabla}^2K}{K^2}$&$=~0$\\ [1ex]
\hline
\vspace{-0.3cm}
& \\
${\mathcal{S}}^{(5)}\equiv U\cdot\left(\frac{\nabla\widetilde{Q}}{\widetilde{Q}}\right)$&$=~0$\\ [1ex]
\hline
\vspace{-0.3cm}
& \\
${\mathcal{S}}^{(6)}\equiv\frac{1}{K} \nabla\cdot\left(\frac{\nabla\widetilde{Q}}{\widetilde{Q}}\right)$&$=~0$\\ [1ex]
\hline
\vspace{-0.3cm}
& \\
${\mathcal{S}}^{(7)}\equiv \frac{K}{D}$&$=~1-\frac{1}{D}$\\ [1ex]
\hline
\vspace{-0.3cm}
& \\
${\mathcal{S}}^{(8)}\equiv \frac{R_{uu}}{K}$&$=~-\lambda$\\ [1ex]
\hline
\hline
\end{tabular}
\end{table} 
\begin{table}[ht!]
\caption{Vector large-$D$ Data in terms of fluid Data} 
\vspace{0.5cm}
\centering 
\label{vector_data}
\begin{tabular}{|c| c|} 
\hline\hline 
 Large-$D$ Data&Corresponding Fluid Data  \\ [1ex] 
\hline 
\hline
\vspace{-0.3cm}
& \\
${V}^{(1)}_{A}~dX^A\equiv P^C_A\left(\frac{\nabla_C K}{K}\right)$ & $=~0$\\ [1ex]
\hline
\vspace{-0.3cm}
& \\
${V}^{(2)}_{A}~dX^A\equiv  P^C_A\left(U\cdot\nabla\right)O_C $&$=~f\left(Q_C\right){\cal P}^{\lambda}_{\mu}\partial_{\lambda}Q_C~dx^{\mu}$\\ [1ex]
\hline
\vspace{-0.3cm}
& \\
${V}^{(3)}_{A}~dX^A\equiv P^C_A\left(U\cdot\nabla\right)U_C$&$=~-f\left(Q_C\right){\cal P}^{\lambda}_{\mu}\partial_{\lambda}Q_C~dx^{\mu}$\\ [1ex]
\hline
\vspace{-0.3cm}
& \\
${V}^{(4)}_{A}~dX^A\equiv P^C_A\left(\frac{\hat{\nabla}^2U_C}{K}\right)$&$=~-f\left(Q_C\right){\cal P}^{\lambda}_{\mu}\partial_{\lambda}Q_C~dx^{\mu}$\\ [1ex]
\hline
\hline
\end{tabular}
\end{table} 
\begin{table}[ht!]
\caption{Tensor large-$D$ Data in terms of fluid Data} 
\vspace{0.5cm}
\centering 
\label{tensor_data}
\begin{tabular}{|c| c|} 
\hline\hline 
 Large-$D$ Data&Corresponding Fluid Data  \\ [1ex] 
\hline 
\hline
\vspace{-0.3cm}
& \\
$t_{AB}\equiv P^C_A P^{C^{\prime}}_B\left[\left(\frac{\nabla_C O_{C^{\prime}}+\nabla_{C^{\prime}} O_{C}}{2}\right)-\frac{P_{CC^{\prime}}}{D}\left(\nabla\cdot O\right)\right]$ & $=~\begin{aligned}
&-r\sigma_{\mu\nu}+\frac{1}{D}\Big(r^2{\cal{P}_{\mu\nu}}+2r\sigma_{\mu\nu}\\
&+rf\left(Q_C\right)\left(u_{\mu}{\cal{P}^{\lambda}_{\nu}}\partial_{\lambda} Q_C+u_{\nu}{\cal{P}^{\lambda}_{\mu}}\partial_{\lambda }Q_C\right)\Big)
\end{aligned}$\\ [1ex]
\hline
\hline
\end{tabular}
\end{table} 
\subsection{Comparing the metrics and gauge fields}\label{sec:metric_match}
In this subsection we shall take the large $D$ limit of the fluid metric and gauge field and match with the metric and gauge field in large $D$ side after expressing them in terms of fluid data. 
\subsubsection{Comparing the gauge fields}
At first we decompose the fluid gauge field into scalar and vector components. As from the gauge condition $A_r$ component of the gauge field is zero to all order in derivative, we write only the components in the boundary directions.
\begin{equation}
\begin{split}
A_{\mu}^{\text{(fluid)}}=\mathcal{B}^{(S)}~u_{\mu}+\mathcal{B}^{(V)}_{\mu}
\end{split}
\end{equation}
where
\begin{equation}
\begin{split}
&\mathcal{B}^{(S)}=\frac{\sqrt{3}~r~Q_C}{2}\left(\frac{r_H}{r}\right)^{D-2}\\
&\mathcal{B}^{(V)}_{\mu}=-2\sqrt{3}\left(\frac{r}{r_H}\right)^D\bigg[1-(D-2)Q_C~f(Q_C)\bigg]F_1^{(1,0)}(\rho,M)~{\cal P}^{\lambda}_{\mu}\left({\partial}_{\lambda}Q_C\right)\\
\end{split}
\end{equation}
The gauge field in large-$D$ side is given as
\begin{equation}
\begin{split}
&A_M=\sqrt{2}\left[\widetilde{f}~O_M+\frac{1}{D}A^{(1)}_M+\mathcal{O}\left(\frac{1}{D}\right)^2\right]\\
&\text{where}~~~\mathcal{A}^{(1)}_M=\mathcal{A}^{(s)}O_M+\mathcal{A}^{(v)}_M
\end{split}
\end{equation}
Also in this case the radial component of the gauge field vanishes and only the components in boundary directions are non-zero. We decompose this into scalar and vector components as follows
\begin{equation}
\begin{split}
A_{\mu}^{\text{(D)}}=\mathcal{Y}^{(S)}~u_{\mu}+\mathcal{Y}^{(V)}_{\mu}
\end{split}
\end{equation}
where
\begin{equation}
\begin{split}
&\mathcal{Y}^{(S)}=-\sqrt{2}~r\left(\widetilde{f}+\frac{1}{D}\mathcal{A}^{(s)}\right)+\mathcal{O}\left(\frac{1}{D}\right)^2\\
&\mathcal{Y}^{(V)}_{\mu}=\frac{1}{D}\sqrt{2}~\mathcal{A}^{(v)}_{\mu}+\mathcal{O}\left(\frac{1}{D}\right)^2\\
\end{split}
\end{equation}
$\mathcal{A}^{(s)}$ is defined as follows
\begin{equation}\label{Ascalar}
\begin{split}
\mathcal{A}^{(s)}&=\sum_{i=1}^{N_s}a^{(i)}_s(Y)~\mathcal{S}^{(i)}\\
&=\left(a^{(2)}_s(Y)~\mathcal{S}^{(2)}+a^{(7)}_s(Y)~\mathcal{S}^{(7)}+K~a^{(R_{uu})}_s(Y)~\mathcal{S}^{(8)}\right)\\
&=\left(-a^{(2)}_s(Y)+a^{(7)}_s(Y)+K~a^{(R_{uu})}_s(Y)\right)\\
&=a^{\text{(total)}}_s(Y)
\end{split}
\end{equation}
where $Y=D(\psi-1)$.\footnote{Here one should note that the integrations in the large-$D$ side are parametrized by $W=D(\psi-1)$. On the other hand we can define another parameter $\widetilde{R}=D\left(\frac{r}{r_H}-1\right)$ to expand other functions in inverse power of dimensions. But it is easy to check that $W=\widetilde{R}+\mathcal{O}\left(\frac{1}{D}\right)$. And hence up to the order we are interested the two parameters are just equal and we simply denote both of them by the $\mathcal{O}(1)$ parameter $Y$ without any further confusion.}\\
where
\begin{equation}\label{atot}
\begin{split}
a^{\text{(total)}}_s(Y)&=-e^{-Y}\left(\frac{1}{N}\right)\int_0^{\infty}~d\rho~e^{-\rho}\int_0^{\rho}d\zeta~e^{\zeta}~s^{\text{(total)}}_{\text{gauge}}(\zeta)+\left(\frac{1}{N}\right)\int_Y^{\infty}d\rho~e^{-\rho}\int_0^{\rho}d\zeta~e^{\zeta}~s^{\text{(total)}}_{\text{gauge}}(\zeta)\\
\end{split}
\end{equation}
And $s^{\text{(total)}}_{\text{gauge}}$ is defined as
\begin{equation}\label{sgtot}
\begin{split}
s^{\text{(total)}}_{\text{gauge}}&=-s^{(2)}_{\text{gauge}}+s^{(7)}_{\text{gauge}}+s^{(R_{uu})}_{\text{gauge}}\\
&=\widetilde{f}\\
&=\widetilde{Q}~e^{-Y}+\mathcal{O}\left(\frac{1}{D}\right)
\end{split}
\end{equation}
Substituting \eqref{sgtot} into \eqref{atot} we have 
\begin{equation}
\begin{split}
&a^{\text{(total)}}_s(Y)=Y~\widetilde{Q}~e^{-Y}
\end{split}
\end{equation}
Putting this in \eqref{Ascalar} we have
\begin{equation}
\begin{split}
\mathcal{A}^{(s)}=Y~\widetilde{Q}~e^{-Y}
\end{split}
\end{equation}
The vector component calculated up to leading order in large-$D$ as
\begin{equation}
\begin{split}
\mathcal{A}^{(v)}_{\mu}&=\sum_{i=1}^{N_s}a^{(i)}_v(Y)~{V}^{(i)}_{\mu}\\
&=f\left(Q_C\right)\left(a^{(2)}_v(Y)-a^{(3)}_v(Y)-a^{(4)}_v(Y)\right){\cal P}^{\lambda}_{\mu}\left({\partial}_{\lambda}Q_C\right)\\
&=\mathcal{O}\left(\frac{1}{D}\right)
\end{split}
\end{equation}
So finally we have
\begin{equation}
\begin{split}
&\mathcal{Y}^{(S)}=-\sqrt{2}~r\left(\widetilde{f}+\frac{1}{D}Y~\widetilde{Q}~e^{-Y}\right)+\mathcal{O}\left(\frac{1}{D}\right)^2\\
&\mathcal{Y}^{(V)}_{\mu}=\mathcal{O}\left(\frac{1}{D}\right)^2\\
\end{split}
\end{equation}
Now we have from \ref{app1} 
\begin{equation}
\begin{split}
F_1^{(1,0)}\left(1+\frac{Y}{D},M\right)=\left(\frac{1}{D}\right)^2
\end{split}
\end{equation}
Now if we expand both large-$D$ and fluid gauge field up to $\mathcal{O}\left(\frac{1}{D}\right)$ by substituting $r=r_H\left(1+\frac{Y}{D}\right)$, we find
\begin{equation}
\begin{split}
&\mathcal{Y}^{(S)}-\mathcal{B}^{(S)}=\mathcal{O}\left(\frac{1}{D}\right)^2\\
&\mathcal{Y}^{(V)}_{\mu}-\mathcal{B}^{(V)}_{\mu}=\mathcal{O}\left(\frac{1}{D}\right)^2\\
\end{split}
\end{equation}
So within the membrane region, the two gauge fields are equivalent to one another.
\subsubsection{Comparing the metric}
As we have discussed earlier we can determine $\mathcal{G}^{\left(rest\right)}_{\mu\nu}$, by simply subtracting the background piece from the full hydrodynamic metric once we know the background metric in $X^A$ coordinates. After decomposing $\mathcal{G}^{\left(rest\right)}_{\mu\nu}$  into scalar, vector and tensor components, we can write it as follows
\begin{equation}
\begin{split}
\mathcal{G}^{\left(rest\right)}_{\mu\nu}&=\mathcal{G}^{(S1)}u_{\mu}u_{\nu}+\mathcal{G}^{(S2)}{\cal P}_{\mu\nu}+\left(\mathcal{G}^{(V)}_{\mu}~u_{\nu}+\mathcal{G}^{(V)}_{\nu}~u_{\mu}\right)+\mathcal{G}^{(T)}_{\mu\nu}\\
\end{split}
\end{equation}
where
\begin{equation}
\begin{split}
&\mathcal{G}^{(S1)}=r^2\bigg(1-V(r)\bigg)\\
&\mathcal{G}^{(S2)}=\mathcal{O}\left(\partial^2\right)\\
&\mathcal{G}^{(V)}_{\mu}=-\frac{3(D-3)~r^2~Q_C}{r_H^2}\bigg[1-(D-2)Q_Cf(Q_C)\bigg]F_1(\rho,M)~{\cal P}^{\lambda}_{\mu}\left({\partial}_{\lambda}Q_C\right)\\
&\mathcal{G}^{(T)}_{\mu\nu}=2r\left(\frac{r}{r_H}F_2\left(\rho,M\right)-1\right)\sigma_{\mu\nu}
\end{split}
\end{equation}
Now we will write the large $D$ metric in terms of fluid data. The large $D$ metric up to first order in $\frac{1}{D}$ can be written as
\begin{equation}
\begin{split}
&{W}_{AB}=\bar{{W}}_{AB}+fO_AO_B+\frac{1}{D}\mathcal{W}^{(1)}_{AB}+\mathcal{O}\left(\frac{1}{D}\right)^2\\
&\text{where}~~~\mathcal{W}^{(1)}_{AB}=\mathcal{Z}^{(s1)}O_AO_B+\left(\mathcal{Z}^{(v)}_AO_B+\mathcal{Z}^{(v)}_BO_A\right)+\mathcal{Z}^{(T)}_{AB}
\end{split}
\end{equation}
Subtracting the background part from this full large-$D$ metric, we will get, ${W}^{\left(rest\right)}_{AB}$. Now from the construction of this metric ${W}^{\left(rest\right)}_{rr}$ and ${W}^{\left(rest\right)}_{r\mu}$ vanishes, and we only have ${W}^{\left(rest\right)}_{\mu\nu}$, which we decompose into scalar, vector and tensor components as follows
\begin{equation}
\begin{split}
{W}^{\left(rest\right)}_{\mu\nu}&=\mathcal{W}^{(S1)}u_{\mu}u_{\nu}+\mathcal{W}^{(S2)}{\cal P}_{\mu\nu}+\left(\mathcal{W}^{(V)}_{\mu}~u_{\nu}+\mathcal{W}^{(V)}_{\nu}~u_{\mu}\right)+\mathcal{W}^{(T)}_{\mu\nu}\\
\end{split}
\end{equation}
where
\begin{equation}
\begin{split}
&\mathcal{W}^{(S1)}=r^2\left(f+\frac{1}{D}\mathcal{Z}^{(s1)}\right)+\mathcal{O}\left(\frac{1}{D}\right)^2\\
&\mathcal{W}^{(S2)}=\mathcal{O}\left(\frac{1}{D}\right)^2\\
&\mathcal{W}^{(V)}_{\mu}=-\frac{1}{D}~r~\mathcal{Z}^{(v)}_{\mu}+\mathcal{O}\left(\frac{1}{D}\right)^2\\
&\mathcal{W}^{(T)}_{\mu\nu}=\frac{1}{D}\mathcal{Z}^{(T)}_{\mu\nu}+\mathcal{O}\left(\frac{1}{D}\right)^2
\end{split}
\end{equation}
So at first we determine $\mathcal{Z}^{(s1)}$ in terms of fluid variables. We write the leading order expressions in large $D$ expansion by substituting $r=r_H\left(1+\frac{Y}{D}\right)$.
\begin{equation}\label{Zs1}
\begin{split}
\mathcal{Z}^{(s1)}&=\sum_{i=1}^{N_s}S^{(i)}_1(Y)~\mathcal{S}^{(i)}\\
&=\left(S^{(2)}_1(Y)~\mathcal{S}^{(2)}+S^{(7)}_1(Y)~\mathcal{S}^{(7)}+K~S^{(R_{uu})}_1(Y)~\mathcal{S}^{(8)}\right)\\
&=\left(-S^{(2)}_1(Y)+S^{(7)}_1(Y)+K~S^{(8)}_1(Y)\right)\\
&=S^{\text{(total)}}_{1s}(Y)
\end{split}
\end{equation}
where
\begin{equation}\label{Stot1s}
\begin{split}
&S^{\text{(total)}}_{1s}(y)=-4\int_y^{\infty}d\rho~\widetilde{f}~a^{\text{(total)}}_s(\rho)-e^{-y}~A_{\text{scalar}}^{\text{(total)}}+\left(\frac{2}{N}\right)\int_y^{\infty}d\rho~e^{-\rho}\int_0^{\rho}d\zeta~e^{\zeta}~s^{\text{(total)}}_{\text{metric}}(\zeta)\\
\text{where}~&A_{\text{scalar}}^{\text{(total)}}=-4\int_0^{\infty}d\rho~\widetilde{f}~a^{\text{(total)}}_s(\rho)+\left(\frac{2}{N}\right)\int_0^{\infty}d\rho~e^{-\rho}\int_0^{\rho}d\zeta~e^{\zeta}~s^{\text{(total)}}_{\text{metric}}(\zeta)\\
\end{split}
\end{equation}
and $s^{\text{(total)}}_{\text{metric}}$ is defined as
\begin{equation}\label{stotmet}
\begin{split}
s^{\text{(total)}}_{\text{metric}}&=-s^{(2)}_{\text{metric}}+s^{(7)}_{\text{metric}}+s^{(R_{uu})}_{\text{metric}}\\
&=-2\widetilde{f}^2\\
&=-2\widetilde{Q}^2~e^{-2~Y}+\mathcal{O}\left(\frac{1}{D}\right)
\end{split}
\end{equation}
Substituting \eqref{stotmet} into \eqref{Stot1s}, we have
\begin{equation}
\begin{split}
&A_{\text{scalar}}^{\text{(total)}}=-3\widetilde{Q}^2\\
&S^{\text{(total)}}_{1s}(Y)=\widetilde{Q}^2e^{-2Y}\left(1-2Y-e^Y\right)
\end{split}
\end{equation}
And hence we have
\begin{equation}
\begin{split}
\mathcal{Z}^{(s1)}&=\widetilde{Q}^2e^{-2Y}\left(1-2Y-e^Y\right)+\mathcal{O}\left(\frac{1}{D}\right)
\end{split}
\end{equation}
Now we will calculate $\mathcal{Z}^{(V)}_{\mu}$ component
\begin{equation}
\begin{split}
 \mathcal{Z}^{(v)}_{\mu}&=f\left(Q_C\right)\left(\mathcal{V}^{(2)}-\mathcal{V}^{(3)}-\mathcal{V}^{(4)}\right){\cal{P}}^{\lambda}_{\mu}\left({\partial}_{\lambda}Q_C\right)\\
 &=\mathcal{O}\left(\frac{1}{D}\right)
\end{split}
\end{equation}
And the tensor component $\mathcal{Z}^{(T)}_{\mu\nu}$ is given by
\begin{equation}
\begin{split}
 \mathcal{Z}^{(T)}_{\mu\nu}&=-2r\left(\frac{D}{K}\right)\text{log}\left(1-\widetilde{Q}^2e^{-Y}\right)\sigma_{\mu\nu}+\mathcal{O}\left(\frac{1}{D}\right)
\end{split}
\end{equation}
So finally we have
\begin{equation}
\begin{split}
&\mathcal{W}^{(S1)}=r^2\left[f+\frac{1}{D}\left(\widetilde{Q}^2e^{-2Y}\left(1-2Y-e^Y\right)\right)\right]+\mathcal{O}\left(\frac{1}{D}\right)^2\\
&\mathcal{W}^{(S2)}=\mathcal{O}\left(\frac{1}{D}\right)^2\\
&\mathcal{W}^{(V)}_{\mu}=\mathcal{O}\left(\frac{1}{D}\right)^2\\
&\mathcal{W}^{(T)}_{\mu\nu}=-r\left(\frac{2}{K}\right)\text{log}\left(1-\widetilde{Q}^2e^{-Y}\right)\sigma_{\mu\nu}+\mathcal{O}\left(\frac{1}{D}\right)^2
\end{split}
\end{equation}
From \ref{app1}, we have
 \begin{equation}
\begin{split}
 F_1\left(1+\frac{Y}{D},M\right)=\mathcal{O}\left(\frac{1}{D}\right)^3
 \end{split}
\end{equation}
hence
\begin{equation}
\begin{split}
\mathcal{G}^{(V)}_{\mu}=\mathcal{O}\left(\frac{1}{D}\right)^2
\end{split}
\end{equation}
and from \ref{appA2}, we have
\begin{equation}
\begin{split}
\mathcal{G}^{(T)}_{\mu\nu}=-r\left(\frac{2}{D}\right)\text{log}\left(1-\widetilde{Q}^2e^{-Y}\right)\sigma_{\mu\nu}+\mathcal{O}\left(\frac{1}{D}\right)^2
\end{split}
\end{equation}
So finally we have
\begin{equation}
\begin{split}
&\mathcal{G}^{(S1)}=r^2\bigg(1-V(r)\bigg)\\
&\mathcal{G}^{(S2)}=\mathcal{O}\left(\partial^2\right)\\
&\mathcal{G}^{(V)}_{\mu}=\mathcal{O}\left(\frac{1}{D}\right)^2\\
&\mathcal{G}^{(T)}_{\mu\nu}=-r\left(\frac{2}{D}\right)\text{log}\left(1-\widetilde{Q}^2e^{-Y}\right)\sigma_{\mu\nu}+\mathcal{O}\left(\frac{1}{D}\right)^2
\end{split}
\end{equation}
So, now if we subtract the fluid metric from the large-$D$ metric both expanded up to $\mathcal{O}\left(\frac{1}{D}\right)^2$, we will find that in the membrane region the two metric matches.
\subsection{Comparing the evolution of two sets of data}\label{sec:eom_match}
In the previous subsection we have seen that the metric and gauge fields of both the perturbation techniques are equivalent in their overlap regime. However, our intention is to show that the solutions generated by these two perturbation techniques are equivalent. Hence we also have to show the equivalence of the differential equations that govern the time evolution of the defining data of both the systems. But the defining data of hydrodynamic metric and large-$D$ metric are constrained by two different looking sets of differential equations. For hydrodynamic metric these equations are given by the equations in \eqref{Hconstraint}, on the other hand the constraint equations for the large-$D$ case are given by equations in \eqref{memceq}.  To show the equivalence of these two gravity solutions we have to show that once `membrane equations' are satisfied, the constraint equations in hydrodynamics are also satisfied. One of these equations is given by\footnote{It is well known that fluid equations can be written as conservation equations of a stress tensor and charge current living on $(D-1)$ dimensional flat space. Also from paper \cite{radiation,Biswas:2019xip} we know that we can define a stress tensor and charge current on the membrane order by order in inverse power of dimensions such that the membrane equations is simply the conservation equations of this stress tensor and charge current. So it would be interesting and easier to show that the conservation of membrane stress tensor and charge current follows from the conservation equations of fluid stress tensor and charge current. But unfortunately at this time we do not have the expressions for the stress tensor and charge current for a charged membrane propagating in AdS space. Hence we are forced to check the equivalence of the two different looking sets of constraint equations, namely membrane equations and fluid equations.}
\begin{equation}\label{memceq1}
P^{A}_C\left[\frac{\widehat{\nabla}^2U_A}{K}-\left(1+\widetilde{Q}^2\right)\frac{\widehat{\nabla}_AK}{K}+U^{B}K_{BA}-\left(1+\widetilde{Q}^2\right)\left(U\cdot\widehat{\nabla}U_A\right)\right]=\mathcal{O}\left(\frac{1}{D}\right)
\end{equation}
We have calculated the different components of these equations in terms of fluid data as follows
\begin{equation}
\begin{split}
&P^{A}_C\frac{\widehat{\nabla}^2U_A}{K}~=~-f\left(Q_C\right){\cal P}^{\lambda}_{\mu}\partial_{\lambda}Q_C\\
&P^{A}_C\frac{\widehat{\nabla}_AK}{K}~=~0\\
&P^{A}_CU^{B}K_{BA}=\mathcal{O}\left(\partial^2\right)\\
&P^{A}_C\left(U\cdot\widehat{\nabla}U_A\right)~=~-f\left(Q_C\right){\cal P}^{\lambda}_{\mu}\partial_{\lambda}Q_C\\
\end{split}
\end{equation}
Substituting these in the L.H.S of equation \eqref{memceq1} evaluate to
\begin{equation}
\begin{split}
&=-f\left(Q_C\right){\cal P}^{\lambda}_{\mu}\partial_{\lambda}Q_C-\left(1+\widetilde{Q}^2\right)\left(-f\left(Q_C\right){\cal P}^{\lambda}_{\mu}\partial_{\lambda}Q_C\right)\\
&=\widetilde{Q}^2~f\left(Q_C\right){\cal P}^{\lambda}_{\mu}\partial_{\lambda}Q_C\\
&=\frac{~3}{8}Q_C^2~f\left(Q_C\right){\cal P}^{\lambda}_{\mu}\partial_{\lambda}Q_C\\
&=\mathcal{O}\left(\frac{1}{D}\right)
\end{split}
\end{equation}
As $f\left(Q_C\right)=\mathcal{O}\left(\frac{1}{D}\right)$, so our membrane velocity and shape function satisfies \eqref{memceq1}.\par 
The second constraint equation is
\begin{equation}
\hat{\nabla}\cdot U=\mathcal{O}\left(\frac{1}{D}\right)
\end{equation}
now $\hat{\nabla}\cdot U=\mathcal{O}\left(\partial^2\right)$. So up to the order we are interested our membrane velocity satisfies this equation.\par 
And the third equation is given by
\begin{equation}\label{memceq3}
\frac{\widehat{\nabla}^2\widetilde{Q}}{K}-U\cdot\widehat{\nabla}\widetilde{Q}-\widetilde{Q}\left[\frac{U\cdot\widehat{\nabla}K}{K}-U\cdot K\cdot U-\frac{R_{uu}}{K}\right]=\mathcal{O}\left(\frac{1}{D}\right)
\end{equation}
The large-$D$ structures appeared in this equation can be calculated in terms of fluid data as
\begin{equation}
 \begin{split}
  &\frac{\widehat{\nabla}^2\widetilde{Q}}{K}=0\\
  &U\cdot\widehat{\nabla}\widetilde{Q}=0\\
  &\frac{U\cdot\widehat{\nabla}K}{K}=0\\
  &U\cdot K\cdot U=-1\\
  &\frac{R_{uu}}{K}=-\lambda
 \end{split}
\end{equation}
Substituting these the L.H.S of \eqref{memceq3}  we can check that \eqref{memceq3} is also satisfied.
So up to the order we are interested our membrane velocity satisfies this equation. So we have shown that our membrane equations follow as a consequence of fluid constraint equations.
\section{Conclusons and future directions :}\label{conclusion}
In this note, we have compared two different perturbation techniques - namely derivative expansion and expansion in inverse powers of dimension, in the regime where both  techniques are applicable.\\
 We have considered the case, when these techniques are used to generate  asymptotically AdS, dynamical black hole type solutions of Einstein-Maxwell systems. We have shown that in the appropriate regime of the parameter space, the two solutions are equivalent to one another upto the first non-trivial order in both the perturbation parameters. It turns out that after a series of gauge transformation and field redefinitions, the metrics and the gauge fields generated by these two different techniques are exactly same upto the order the solutions are known on both sides. \\
This work could be extended to many directions. Below we are listing few of them.\\
 
As mentioned before, this equivalence is very much expected on physical grounds. Still we believe it is important to chart out the subtle details of the redefinitions and transformation it involves.
If we know  how and when the two perturbation techniques generate the same solution, it will help us to find out when they are really different where one is generating new set of dynamical black hole solutions that could not be generated from the other. So once we identified and studied the overlapping regime of the parameter space, it would be interesting to look at the non-overlapping regimes.\par
Because of the very generic nature of the physical intuition that asserts this equivalence, 
we believe that it exists not only for pure Einstein systems or Einstein-Maxwell systems but also for Einstein-dilaton systems, higher-derivative gravity theory \cite{Bhattacharyya:2008ji,Armas:2016mes,Chen:2017rxa,articleDutta}
 or any other systems where we can apply both the perturbation techniques. However, our explicit calculations are very much system-specific, which somehow obscures this genericity. It would be interesting to set these calculations more physically or abstractly, without using too much details of a given theory.\par
In some sense, this note is also describing a duality between  the dynamics of a $(D-1)$ dimensional  charged and  massive membrane, embedded in $D$ dimensional AdS space  and  that of charged fluid, living on the boundary of the AdS space. The basic variables of charged fluids are temperature ($T$), velocity ($u^\mu$) and charge density ($Q$); whereas any charged dynamical membrane would be characterized by the embedding function ($\psi$),  the charge density field ($\tilde Q$) and the velocity field ($U^A$)\footnote{Just like the velocity field in fluids, the membrane velocity $U^A$  captures the charge or mass redistribution within the membrane.}. The statement of the duality could be as follows.\\
If it is possible to have an all order completion for both the membrane equations and the relativistic fluid equations, then they  are actually the same equations, just written in terms of two different set of variables.\\
 In this note, we have worked out this variable redefinition (see equations \eqref{psi_redefinition},\eqref{O_redefinition} and \eqref{U_redefinition}), needed to show the equivalence, upto the order the equations are known on both sides.\\
 As it is clear from all previous discussions, the key reason for this equivalence is simply the fact that both the systems are dual to the same gravity solution in the overlap regime of the two perturbation parameters. Still it might be possible to formulate the duality, removing the gravity altogether, because once we know the appropriate variable redefinition, that is enough to show the equivalence of the membrane and fluid equations. However,  we should emphasize that at  this stage it is a duality between a very specific membrane and a very specific fluid, the ones that could have gravity duals. It would be really interesting to see if we can extend such duality to more generic fluids and membranes. It will provide new ways to analyze the unsolved problems in both sets of equations.\\
\section*{Acknowledgement}
I am extremely grateful to Sayantani Bhattacharyya for suggesting me this problem and providing guidance at every stage of this project. I am also thankful to her for going through the draft of this note several times and for very useful suggestions. I would like to thank Parthajit Biswas and Anirban Dinda for collaboration at the initial stage of this project. I would also like to thank Parthajit Biswas, Anirban Dinda, Suman Kundu, Sarthak Duary and Shuvayu Roy for useful discussion. Finally, I would like to acknowledge my gratitude to the people of India for their support to the research in basic sciences.
\appendix
 \section{The Large-$D$ limit of the integrations appearing  in hydrodynamic metric}\label{app:largeD}
In this section we will try to do the integrations appearing in the fluid metric and we will do those integrals in $\frac{1}{D}$ expansion.
\subsection{Analysis of the integral in the function $F_1$}\label{app1}
The integral appearing in $F_1$ is given by
\begin{equation}
\begin{split}
F_3(\rho,M)&=\int_{\rho}^{\infty}dp\frac{1}{\left(1+\frac{1}{4}\frac{3(D-3)}{2(D-2)}\frac{Q_1^2}{p^{2(D-2)}}-\frac{M}{p^{D-1}}\right)^2}\Big(\frac{1}{p^{2(D-1)}}-\frac{c_1}{p^{2D-3}}\Big)\\
&=\int_{\rho}^{\infty}dp\frac{1}{\left(1+\frac{\left(M-1\right)}{p^{2(D-2)}}-\frac{M}{p^{D-1}}\right)^2}\Big(\frac{1}{p^{2(D-1)}}-\frac{c_1}{p^{2D-3}}\Big)
\end{split}
\end{equation}
Though it is difficult to do the integral for general $M$, it is easier to do the integral for $M=1$. Hence we have done the integration by expanding $M$ around $1$ as $M=1+\delta M$ in a power series of $\delta M$. It is easy to do the integrals appearing in  each coefficient in the power series of $\delta M$ and if we do those integrals and expand each of the results of those integrals in $\frac{1}{D}$ as $\rho=1+\frac{Y}{D}$ we will find that
\begin{equation}
\begin{split}
F_3(1+\frac{Y}{D},M)=\left(\frac{1}{D}\right)^2
\end{split}
\end{equation}
\subsection{Analysis of the integral in the function $F_2$}\label{appA2}
The integral appearing in $F_2$ is given by
\begin{equation}
\begin{split}
&F_2(\rho,M)=\int_{\rho}^{\infty}\frac{p^D\left(p^D-p^2\right)}{p^2\left(p^{2D}-M p^{D+1}+(M-1)p^4\right)}dp\\
\end{split}
\end{equation}

This integration is also difficult to do other than $M=1$ and hence by pursuing the previous method at first we expand the integrand around $M=1$ by putting $M=1+\delta M$. Then the integrand becomes
\begin{equation}
\begin{split}
\frac{p^D\left(p^D-p^2\right)}{p^2\left(p^{2D}-M p^{D+1}+(M-1)p^4\right)}&=\frac{p^{D-2}-1}{p\left(p^{D-1}-1\right)}+\frac{\left(p^D-p^3\right)\left(p^D-p^2\right)}{p^{D+1}\left(p^D-p\right)^2}{\delta M}\\
&+\frac{\left(p^D-p^3\right)^2\left(p^D-p^2\right)}{p^{2D}\left(p^D-p\right)^3}{\delta M}^2+\cdots
\end{split}
\end{equation}

If we do each of these integrals for each of the coefficients of $\delta M,~{\delta M}^2, \cdots$ and then expand those results in $\frac{1}{D}$ by replacing $\rho=1+\frac{Y}{D}$, we will get terms like $e^{-Y}\left(\frac{\delta M}{D}\right)$, $e^{-2Y}\left(\frac{{\delta M}^2}{2D}\right), e^{-3Y}\left(\frac{{\delta M}^3}{3D}\right), \cdots$. If we combine all those terms for power of $\delta M$ $\geq 1$ we will get $-\frac{1}{D}\text{log}[1-\widetilde{Q}^2e^{-Y}]$ and for the $\text{0}^{th}$ order term in $\delta M$ following the appendix A of \cite{Bhattacharyya2019leading}  we have $1+\left(\frac{1}{D}\right)^2$. 
\section{The inverse of the background metric and christoffel symbols w.r.t background metric}\label{bacground_christoffel}
We can find out the inverse of the background metric order by order in derivative expansion. These expressions are given by
\begin{equation}
\begin{split}
\bar{\mathcal{G}}^{rr}&=r^2-2r\frac{\Theta}{D-2}+\mathcal{O}\left(\partial^2\right)\\
\bar{\mathcal{G}}^{\mu r}&=u^{\mu}-\frac{a^{\mu}}{r}+\mathcal{O}\left(\partial^2\right)\\
\bar{\mathcal{G}}^{\mu\nu}&=\frac{1}{r^2}{\cal P}^{\mu\nu}-\frac{2}{r^3}\sigma^{\mu\nu}+\mathcal{O}\left(\partial^2\right)\\
\end{split}
\end{equation}
Having these expression for the background metric we have the christoffel symbols for the background metric as
 \begin{equation}
\begin{split}
&\bar{\Gamma}^r_{rr}=0\\
&\bar{\Gamma}^{\alpha}_{rr}=0\\
&\bar{\Gamma}^r_{\mu r}=r~u_{\mu}-\frac{\Theta}{D-2}u_{\mu}\\
&\bar{\Gamma}^{\alpha}_{\mu r}=\frac{1}{r}{\cal P}^{\alpha}_{\mu}-\frac{1}{2~r^2}\left(u_{\mu}a^{\alpha}-a_{\mu}u^{\alpha}\right)+\frac{1}{2~r^2}\left(\partial^{\alpha}u_{\mu}-\partial_{\mu}u^{\alpha}\right)-\frac{1}{r^2}\sigma^{\alpha}_{\mu}\\
&\bar{\Gamma}^r_{\alpha\mu}=-r^3\left({\cal P}_{\alpha\mu}-u_{\alpha}u_{\mu}\right)+{r^2}\left(u_{\alpha}a_{\mu}+u_{\mu}a_{\alpha}\right)-2~r^2\sigma_{\alpha\mu}+r^2\frac{\Theta}{D-2}{\cal P}_{\alpha\mu}-3r^2\frac{\Theta}{D-2}u_{\alpha}u_{\mu}\\
&\bar{\Gamma}^{\beta}_{\alpha\mu}=u^{\beta}\left(-r\left({\cal P}_{\alpha\mu}-u_{\alpha}u_{\mu}\right)+\left(u_{\alpha}a_{\mu}+u_{\mu}a_{\alpha}\right)-2\sigma_{\alpha\mu}-\frac{\Theta}{D-2}u_{\alpha}u_{\mu}-\frac{\Theta}{D-2}{\cal P}_{\alpha\mu}\right)+a^{\beta}\left({\cal P}_{\alpha\mu}-u_{\alpha}u_{\mu}\right)\\
\end{split}
\end{equation}
\section{Notation}
In this section we write down the various symbols we have used throughout this note.
\begin{align*}
\bar{G}_{AB}~&:~\text{Background metric in }Y^A\equiv\{\rho,y^{\mu}\}~\text{coordinates}\\
\bar{\mathcal{G}}_{AB}~&:~\text{Background metric in }X^A\equiv\{r,x^{\mu}\}~\text{coordinates}\\
{\mathcal{G}}_{AB}~&:~\text{Full metric in }X^A\equiv\{r,x^{\mu}\}~\text{coordinates}\\
\bar{W}_{AB}~&:~\text{Background metric in arbitrary coordinates}\\
{W}_{AB}~&:~\text{Full metric in arbitrary coordinates}\\
{\eta}_{\mu\nu}~&:~\text{The boundary metric}\\
{u}_{\mu}~&:~\text{Fluid velocity}\\
{U}_{A}~&:~\text{Membrane velocity}\\
Q~&:~\text{Charge field in hydrodynamic metric}\\
\widetilde{Q}~&:~\text{Charge field in large-$D$ metric}\\
\Pi_{AB}~&:~\text{Projector perpendicular to $n_A$}\\
P_{AB}~&:~\text{Projector perpendicular to both $n_A$ and $U_A$}\\
{\cal P}_{\mu\nu}~&:~\text{Projector perpendicular to $u_{\mu}$}\\
\nabla_{A}~&:~\text{Covarient derivative w.r.t background metric}\\
\hat{\nabla}~&:~\text{Covarient derivative projected along the membrane}\\
\end{align*}

 \bibliographystyle{JHEP}
\bibliography{larged3}

\end{document}